\documentclass[aps,physrev,reprint,groupedaddress]{revtex4-2}

\usepackage{graphicx}
\usepackage{dcolumn}
\usepackage{bm}
\usepackage{hyperref}
\usepackage{xcolor}
\usepackage{amsmath}
\usepackage{comment}

\begin{document}

\title{A theory for coexistence and selection
of branched actin networks\\
in a shared and finite pool of monomers} 

\author{Valentin Wössner}

\author{Falko Ziebert}

\author{Ulrich S. Schwarz}
 \email{schwarz@thphys.uni-heidelberg.de}

 \affiliation{Institute for Theoretical Physics, Heidelberg University, Philosophenweg 19, 69120 Heidelberg, Germany}
 \affiliation{BioQuant, Heidelberg University, Im Neuenheimer Feld 267, 69120 Heidelberg, Germany}

\date{\today}

\begin{abstract}
Cellular actin structures are continuously turned over while keeping similar sizes. Since they all compete for a shared pool of actin monomers, the question arises how they can coexist in these dynamic steady states. Recently, the coexistence of branched actin networks with different densities growing 
in a shared and finite pool of purified proteins has been demonstrated in a biomimetic bead assay.
However, theoretical work in the context of organelle size regulation has mainly been focused on linear architectures, such as single filaments and bundles, and thus is not able to explain this observation.
Here we show theoretically that the local depletion of actin monomers caused by the growth of a
branched network naturally gives rise to a negative feedback loop between network density and growth rate, and that this competition is captured by one central \textcolor{black}{ordinary differential} equation. A comprehensive bifurcation analysis shows that
the theory leads to well-defined steady states even in the case of multiple networks sharing the same pool of monomers, without any need for specific molecular processes. Under increasing competition strength,
coexistence is replaced by selection. 
We also show that our theory is in excellent agreement with spatiotemporal simulations, implemented in a finite element framework,  
\textcolor{black}{and that local depletion even occurs in the presence of a large pool of non-polymerizable actin.}
In summary, our work suggests that local monomer depletion is the decisive 
and universal factor controlling growth of branched actin networks.
\end{abstract}

\maketitle


\section{Introduction} \label{sec: Introduction}

The actin cytoskeleton is hierarchically organized, where actin monomers, as smallest subunits, assemble into linear filaments which in turn form higher-order structures \cite{blanchoinActinDynamicsArchitecture2014}. This assembly is orchestrated by a variety of actin binding proteins, resulting in different geometries such as linear bundles or branched networks \cite{lappalainenBiochemicalMechanicalRegulation2022,goodeMechanismsActinDisassembly2023}. These structures are crucial for essential cellular processes such as division, migration, and shape regulation. Therefore, they must be stable to ensure proper functioning, while also being quickly adaptable to changing environmental conditions \cite{banerjeeActinCytoskeletonActive2020}. This duality of stability and adaptability is enabled by so-called dynamic steady states, in which monomers are constantly turned over under the consumption of energy, while size and geometry remain the same \cite{theryReconstitutingDynamicSteady2024}. This balance can either be local, where binding and unbinding rates are equal at the same position, or it can be due to global fluxes, like in treadmilling, which has been demonstrated both for single filaments \cite{wegnerHeadTailPolymerization1976} and whole lamellipodial structures \cite{carlierGlobalTreadmillingCoordinates2017}.

Size control in the dynamic systems of the cell is a fundamental and very important
question and various mechanisms have been proposed \cite{chanHowCellsKnow2012,banerjeeDesignPrinciplesFeedback2025}. So-called limited pool models \cite{goehringOrganelleGrowthControl2012} present an elegant solution because they do not involve additional proteins, which alter molecular rates in a size- or age-dependent way. Instead, a finite pool of polymerizable monomers gets depleted by a growing structure until the critical assembly concentration is reached from above. For a single structure, this results in a well-defined steady state, subject only to small stochastic fluctuations. \textit{In vivo} studies indeed indicate that the amount of available actin is limiting growth. However, they also show that most, if not all, cytoplasmic actin structures are sharing the same pool \cite{rottyCompetitionCollaborationDifferent2015, kadzikFActinCytoskeletonNetwork2020} and that diffusion of monomers is global \cite{raz-benaroushActinTurnoverLamellipodial2017}. This poses conceptual problems if 
more than one structure is growing in this shared pool \cite{mohapatraLimitingPoolMechanismFails2017}. For identical structures, assembling at the same rates, the obtained system of equations is underdetermined, resulting in large fluctuations. Even worse, if one structure has a larger critical concentration than another, it will shrink to zero size because the pool is depleted too much.

To address these issues, the limiting pool model has been extended
by size-sensitive mechanisms. Banerjee and Banerjee presented a general theory for size-dependent growth in a finite pool, where both assembly and disassembly rates have a power law feedback \cite{banerjeeSizeRegulationMultiple2022}. They showed that negative feedback, either via a decreasing assembly or via increasing disassembly, or via a combination of both, 
leads to coexistence in a broad parameter regime. The actual nature
of the feedback mechanism and the exponents of the power laws depend on the details
of a specific system. Thus, their theory is universal in the sense that “size” can be anything, e.g. the length of a filament, the volume of an organelle or the density of a network.

For the actin system, which is the most
prominent example of a limited pool system, 
detailed models for such mechanisms have mainly considered linear filaments \cite{mohapatraDesignPrinciplesLength2016}, whose dynamics are very well described by master equations. More recently, the growth of actin cables has been investigated, for which the crosslinking of several filaments into a bundle can explain the observed length distributions in the absence of a direct size-dependent feedback \cite{rosarioUniversalLengthFluctuations2023, mcinallyLengthControlEmerges2024} and the coexistence in a shared pool with severing \cite{mominQuantitativeSignaturesDisassembly2025}. Thus, size control might naturally emerge from geometric considerations in higher-order structures \cite{mcinallyLengthControlEmerges2024}.
However, these explanations cannot be easily transferred to structures that are not linear bundles.

Recently, size control and coexistence have been demonstrated experimentally 
for branched actin structures
for the first time, namely in a biomimetic bead assay with a minimal set of proteins \cite{guerinBalancingLimitedResources2025}. By placing two or more beads with well-defined surface-coverages of 
nucleation promoting factors (NPFs)
into a microwell, containing the reaction mix for which the amount of all proteins is precisely known, the simultaneous growth of several networks could be observed.
They studied two different setups. 
In the first one, the surface density 
of NPFs was the same on all the beads, i.e. the assembly properties of all networks were the same. Then, all networks grew to the same size, demonstrating proper size control. When more beads were added into one micro\-well, the individual network sizes became smaller, suggesting that the monomer pool is actually limiting. In the second setup, some beads were coated with only half the NPF-density, such that their growth rate is diminished. Then, a network still formed at this “weak” bead as long as the competition was not too strong, meaning that only a few “strong” beads with the higher coating density, 
were present. The more strong beads were present, the less likely it was that the weak network was growing, until it eventually became impossible, demonstrating the limits of coexistence and a crossover to selection
of the fittest.

To theoretically address these experimental observations, one has to go beyond linear structures and combine
existing concepts for limiting pools with the known physics of growing
actin networks. It is well known that the main processes in growing actin
networks are Arp2/3-mediated branching and stop of filament growth
by capping proteins. 
\textcolor{black}{Several modeling frameworks have been devised to describe how different
processes work together in branching actin networks. Agent-based computer simulations
with software packages such as
Cytosim, MEDYAN or AFINES
have been used mainly in the context
of clathrin-mediated endocytosis and
growth of lamellipodia
\cite{mundSystematicNanoscaleAnalysis2018, akamatsuPrinciplesSelforganizationLoad2020, popovMEDYANMechanochemicalSimulations2016, limanRoleArp232020, freedmanVersatileFrameworkSimulating2017, leeNewProposedMechanism2008}.
While they can be used to predict
the macroscopic effect of microscopic
processes such as branching, they require a lot of 
computing time and cannot be easily 
generalized. Two-dimensional network
simulations have been used to 
explain transitions between
network architectures with different 
orientation patterns and to validate simple rate equations for the transitions \cite{malySelforganizationPropulsiveActin2001, weichselMesoscopicModelFilament2013,muellerLoadAdaptationLamellipodial2017, garnerLeadingEdgeMaintenance2022}.
An intermediate level of modeling
are continuum models with spatial
dimensions, usually formulated
as partial differential equations (PDEs), which have been used
to describe the turnover in the
lamellipodium of migrating cells \cite{mogilnerRegulationActinDynamics2002, raz-benaroushActinTurnoverLamellipodial2017} or the growth speed against movable or deformable objects \cite{boukellalSoftListeriaActinbased2004, paluchDeformationsActinComets2006}.} Lattice models, incorporating aging and severing of filaments, can predict the length and width of actin networks \cite{carlssonDisassemblyActinNetworks2007, michalskiEffectsFilamentAging2010, michalskiModelActinComet2011, manhartQuantitativeRegulationDynamic2019}. The effect of diffusion-limited network growth was investigated for “lamellipodium-like” structures \cite{boujemaa-paterskiNetworkHeterogeneityRegulates2017} as well as for bead motility assays \cite{paluchDeformationsActinComets2006}.
However, none of these studies considered local protein depletion in the context of co-assembling structures in a limited pool and the possible implications for coexistence.

To bridge this currently existing gap between network and limited pool models, here we propose local depletion of actin monomers at the leading edge of a dense network as a naturally arising negative feedback between the current state of a structure (filament density) and its growth rate (creation of new branches),
and therefore as a mechanism for the coexistence of networks observed in the bead motility assay \cite{guerinBalancingLimitedResources2025}. 
\textcolor{black}{Motivated by the success of earlier work on branching actin
networks using rate equations \cite{malySelforganizationPropulsiveActin2001, carlssonGrowthVelocitiesBranched2003, weichselTwoCompetingOrientation2010, mullinsSolutionSurfaceFilament2018}, here we 
present a theory centered around 
an ordinary differential equation (ODE)
whose steady states can be mathematically analyzed
with bifurcation theory.}
We show that this feedback has the same form as suggested in the general framework for coexistence \cite{banerjeeSizeRegulationMultiple2022} and derive a general model of coupled ODEs combining local depletion in a limited pool with the basic assembly properties of branched networks. 
The steady state problem can be expressed in terms of the network densities at the leading edge, reducing the number of equations to the number of inequivalent networks in competition. This allows us to reproduce the experimental results of Guérin et al. \cite{guerinBalancingLimitedResources2025}. We also present a phase diagram for the transition from coexistence to selection under increased competition. 

This paper is structured as follows. In section \ref{sec: Model}, we first recall the 
underlying mechanisms of actin network growth and on this basis then
derive the governing equations of our theory in three steps. First, we calculate an analytical expression for the local depletion in the case of a single network. Then, we estimate the steady state length of the network based on the loss of percolation. Finally, we generalize to multiple networks, which effectively interact through the shared actin pool. We derive one central equation, which is then analyzed in section \ref{sec: Results}. By first considering a single network, we estimated the two unknown parameters in our model by comparison to the experimental measurements. We then present the results for the case of two or more identical networks, and finally demonstrate coexistence of networks with different assembly rates. The analytical results are verified by solving the full three-dimensional \textcolor{black}{diffusion} problem numerically.
\textcolor{black}{Within the numerical framework, we furthermore show that even in the presence of a buffer protein like thymosin, which sequesters large amounts of actin in a non-polymerizable state acting as a reservoir, as observed \textit{in vivo} \cite{cassimerisThymosinBeta41992, goldsteinThymosinV4Actinsequestering2005, raz-benaroushActinTurnoverLamellipodial2017}, the assembly can still be limited by diffusion.}
In the last section \ref{sec: Discussion}, we discuss the implications of our results for the understanding of the co-assembly of actin structures and the role of protein depletion.

\section{Model} \label{sec: Model}

\begin{figure*}[t]
    \centering
    \includegraphics[width=\textwidth]{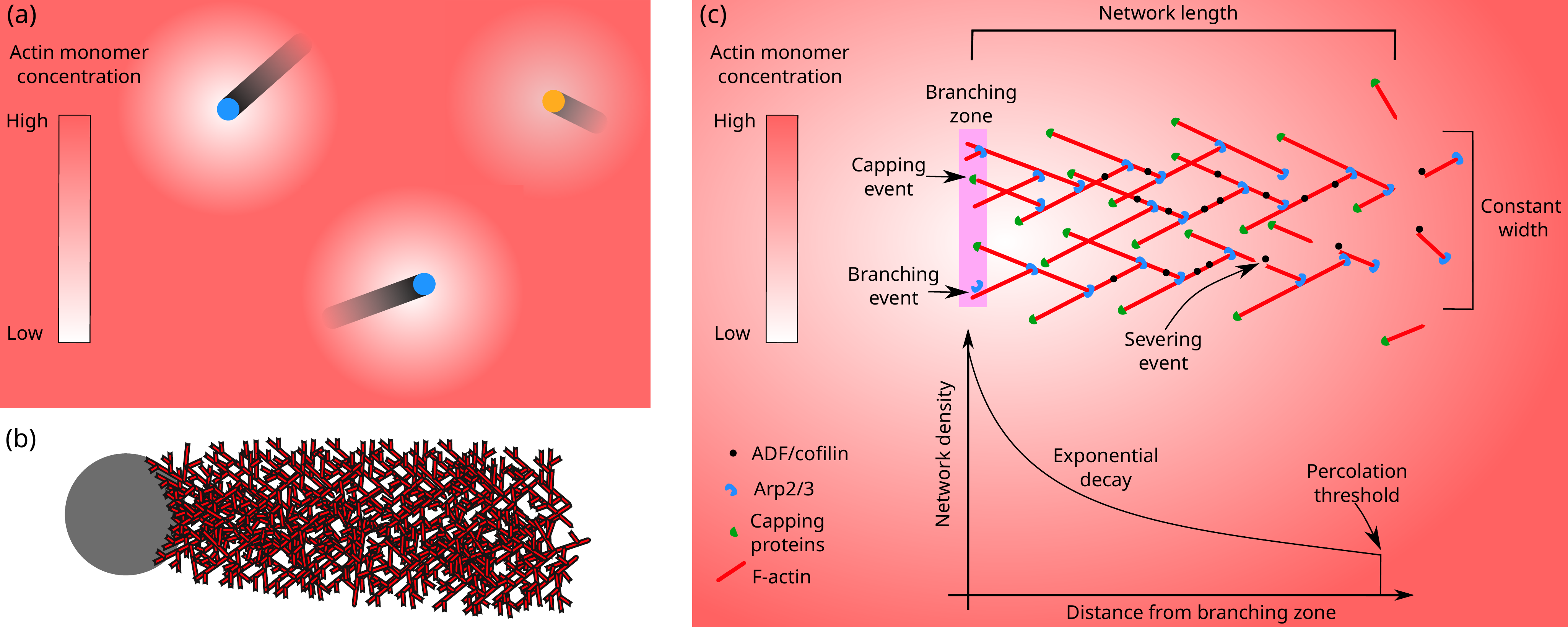}
    \caption{Experimental situation of interest. (a) Actin networks grown from beads with high (blue) and low (orange) coating densities 
    of nucleation promoting factor coexist in a shared pool of actin monomers. The “strong” beads create denser networks (black tails) than the “weak” beads (grey tails). They consume more monomers per time and thus deplete the local actin monomer concentration to a higher degree, indicated by the color gradient from red (high actin monomer concentration) to white (low concentration). Even though all beads share one pool of monomers, the local concentration of actin monomers is higher at the weak bead. (b) Sketch of an actin network growing from the bead surface.
    (c) Schematic sketch of the mechanisms underlying network growth: Arp2/3 is
    activated at the surface and leads to a thin branching zone. There, the balance of branching and capping determines the filament density of the network. The network grows in length by filament elongation at the front and is disassembled by ADF/cofilin-mediated severing at the back, leading to an exponential density profile along its length with a constant network width. The position at which the density drops below the percolation threshold defines the network size. Branching and filament elongation speed are limited by the local concentration of actin monomers, which are locally depleted in the branching zone and are replenished by diffusive fluxes.}
    \label{fig: model sketch}
\end{figure*}

\subsection{Molecular mechanisms underlying network growth}
\label{sec: Molecular mechanisms underlying network growth}

Reconstituted systems with a minimal set of purified proteins are a powerful tool for quantitative measurements. To study branched actin networks, formed by the Arp2/3 complex, bead motility assays are a well-established method, mimicking the movement of intracellular pathogens like Listeria monocytogenes \cite{loiselReconstitutionActinbasedMotility1999}. Polystyrene beads are coated with NPFs. In a solution containing actin monomers and the Arp2/3 complex, NPFs activate the latter and start the formation of a branched network. Since NPFs are surface-bound, activated Arp2/3 is only present in a thin region of a few hundred nanometers around the bead \cite{wearActinDynamicsAssembly2000}, which is much smaller than the typical network length. Since all branching activity happens there, we call this the \textit{branching zone}. Furthermore, Arp2/3 branching is autocatalytic \cite{lappalainenBiochemicalMechanicalRegulation2022} meaning that activated Arp2/3 attaches to already existing filaments and creates a branch point with the characteristic angle of 70$^\circ$ between the mother and daughter filament. After release from the NPF, the Arp2/3 is integrated into the network and the exposed barbed end of the daughter filament starts to grow if the monomer concentration in solution is higher than the critical concentration of 0.12 $\mu$M \cite{pollardRateConstantsReactions1986}. The geometry of the network, i.e. the orientation of filaments relative to the overall network direction, varies depending on actin binding protein concentrations and loading conditions \cite{weichselTwoCompetingOrientation2010, bielingForceFeedbackControls2016}. Due to the very well-defined branching angle of 70$^\circ$ of Arp2/3, there are two idealized patterns possible \textcolor{black}{in the lamellipodia of cells migrating on flat (2D)
surfaces}: either $\pm35^\circ$ or $0,\pm70^\circ$ orientations \cite{malySelforganizationPropulsiveActin2001,weichselTwoCompetingOrientation2010,muellerLoadAdaptationLamellipodial2017,svitkinaUltrastructureActinCytoskeleton2018}.

In addition to Arp2/3, regulatory proteins are necessary for efficient network formation and bead propulsion. Capping proteins bind to barbed ends and will arrest elongation, preventing so-called “fishbone” patterns \cite{pantaloniArp23Complex2000, wiesnerBiomimeticMotilityAssay2003}. Profilin forms a complex with actin monomers and facilitates ADP-ATP exchange and barbed end polymerization. 
\textcolor{black}{Another actin monomer-binding protein, thymosin,
can bring them into a non-polymerizable state
and adds an additional level of control on their
availability for polymerization
\cite{cassimerisThymosinBeta41992, goldsteinThymosinV4Actinsequestering2005}.}
In the absence of any disassembly factors, the networks are stable for several hours or even days \cite{colinRecyclingActinMonomer2023, guerinBalancingLimitedResources2025}, indicating that simple monomer depolymerization at pointed ends is not sufficient to break down the network, which is most likely also the case \textit{in vivo} \cite{raz-benaroushActinTurnoverLamellipodial2017}. \textcolor{black}{Only after adding e.g. actin depolymerization factor cofilin (ADF/cofilin), turnover
of the actin in the network occurs on the minute timescale.}
ADF/cofilin binds to actin filaments, primarily to ADP-actin, and significantly increases the severing rate \cite{michelotActinFilamentStochasticDynamics2007}, enhances depolymerization at pointed ends \cite{pavlovActinFilamentSevering2007} and facilitates debranching \cite{chanCofilinDissociatesArp22009}. 
Like profilin, cyclase-associated protein recharges actin monomers, released from the network, with ATP and interacts synergistically with ADF/cofilin accelerating overall turnover. This creates a stable dynamic steady state of a globally treadmilling system, pushing the bead forwards.
In this steady state, the bead velocity and the network size are constant over several hours in a closed system with finite resources \cite{colinRecyclingActinMonomer2023}.

It has been observed in reconstituted systems that fast monomer consumption of dense networks can locally deplete monomer concentration and slow down growth such that the system becomes diffusion-limited \cite{plastinoEffectDiffusionDepolymerization2004, boujemaa-paterskiNetworkHeterogeneityRegulates2017, bleicherDynamicsActinNetwork2020}. This in turn can steer the direction of growth in heterogeneously dense networks \cite{boujemaa-paterskiNetworkHeterogeneityRegulates2017}. On the other hand, depletion of depolymerization factors like ADF/cofilin is also possible such that density and width display a positive feedback and increase network length \cite{manhartQuantitativeRegulationDynamic2019}. Even in the absence of any depletion effects, a network subject to severing is more stable the denser it is \cite{boujemaa-paterskiNetworkHeterogeneityRegulates2017} and the network length is inversely proportional to its mesh size \cite{michalskiModelActinComet2011}. These observations lead us to the conclusion that the filament density of a network is the main dynamic quantity to consider in order to understand network stability and, in turn, the coexistence observed by Guérin et al. \cite{guerinBalancingLimitedResources2025}.

\subsection{Growth equation}

In the recent bead motility assay
\cite{guerinBalancingLimitedResources2025},
it has been 
shown that different actin networks
can coexist in steady state equilibria,
even if they differ in size due to
different conditions for filament
nucleation, as shown schematically
in Fig. \ref{fig: model sketch}(a).
In the experiment, the whole bead surface is coated with NPFs, but after some lag time, this symmetry is spontaneously broken, with the network forming predominantly on one side of the bead \cite{vanoudenaardenCooperativeSymmetrybreakingActin1999}, see Fig. \ref{fig: model sketch}(b), leading to a stable “comet tail” and directed movement. Here we ignore the curved bead geometry and simply consider a network growing from a flat surface of size $\pi R^2$ with the bead radius $R$, i.e. we assume a cylindrical network with the same radius as the bead.
Fig. \ref{fig: model sketch}(c) illustrates
how the size of this network is controlled
by the competition of network assembly
on the bead surface and network disassembly
towards the end of the comet tail.
\textcolor{black}{In the growth zone, 
capped filaments do not elongate anymore and are quickly pushed backwards. Because they cannot contribute to the branching process, only uncapped filaments are usually considered as dynamic quantities \cite{malySelforganizationPropulsiveActin2001, carlssonGrowthVelocitiesBranched2003, weichselTwoCompetingOrientation2010, mullinsSolutionSurfaceFilament2018}. Indeed it has been confirmed in stochastic network simulations
that the uncapped filaments dominate the dynamics
in the branching zone \cite{weichselTwoCompetingOrientation2010, weichselMesoscopicModelFilament2013, weichselUnifyingAutocatalyticZerothorder2013}. 
For the total number of uncapped filaments within this region, $N$, we describe the change in its area density, $n = N / (\pi R^2)$, by the simple mass action equation
}
\begin{equation} \label{eq: filament density ODE}
    \dot{n} = \left ([\text{NPF}] [\text{Arp}] k_{\text{branch}} g(n) - [\text{cap}] k_{\text{cap}} \right) n.
\end{equation}
Here we have introduced the surface density of NPFs, $[\text{NPF}]$, the volume concentrations of Arp2/3 and capping proteins, $[\text{Arp}]$ and $[\text{cap}]$, respectively, and the corresponding effective rates of branching and capping, $k_{\text{branch}}$ and $k_{\text{cap}}$. The concentration of actin monomers in the branching zone, $g(n)$, depends on the current state because of local and global depletion of monomers.
The goal of the remainder of this section is to determine $g(n)$.
However, already at this stage we can see
that two steady states of
this system exist, namely  $n=0$ and $g(n) = ([\text{cap}] k_{\text{cap}})/([\text{NPF}] [\text{Arp}] k_{\text{branch}})$.
\textcolor{black}{
In Appendix \ref{sec: Spatial model of actin network}, we show how Eq.~\eqref{eq: filament density ODE} can be motivated from a more detailed one-dimensional PDE-model that considers both capped and uncapped filament densities along the actin tail.}

\subsection{Local monomer depletion as negative feedback}
\label{sec: Local monomer depletion as negative feedback}

Local depletion occurs when the available monomers in the branching zone are consumed faster by the growing network than new monomers are replenished by diffusion from the surrounding reservoir. This has been observed for growing actin networks in reconstituted systems and poses a limiting factor to the growth speed \cite{boujemaa-paterskiNetworkHeterogeneityRegulates2017, bleicherDynamicsActinNetwork2020}. The elongation of an actin filament is simply given by the constant polymerization rate $k_{\text{poly}}$ and the concentration of actin monomers at that position $g(n)$. At the considered concentrations of actin monomers of a few $\mu$M, the rate of unbinding at the barbed ends is negligible compared to the speed of binding \cite{mogilnerRegulationActinDynamics2002}. Therefore, the total number of monomers getting polymerized by the growing network per second is given by $\pi R^2 k_{\text{poly}} g(n) n$. Since the bead is in solution, monomers will be replenished from all directions and the problem can be described by the diffusion equation with a sink at the bead position. For simplicity, we first assume a spherical symmetry, neglecting the influence of the bead and of the actin network on the monomer diffusion. Furthermore, we model the monomer consumption in the branching zone as a stationary, spherical sink of radius $R$ (i.e. non-point like) at the origin to retain the symmetry. Instead of an explicit source term describing the monomers getting released from the network in the disassembly process, we use the fact that the monomer concentration integrated over the volume of the microwell, $V$, should be equal to the total amount of monomers left in solution, $G(n)$. This will lead to an influx, i.e. non-vanishing derivative, at the outer boundary to fulfill the solvability condition of the Laplace equation.
The steady state problem is then described by a one-dimensional boundary value problem along the radial component $r$ with one boundary condition and an additional integral constraint
\begin{subequations}
\begin{align}
    \frac{D}{r^2} \partial_r \left( r^2 \partial_r g(n,r) \right) &= 0, \label{eq: diffusion PDE SS} \\
    \int_V g(n,r) \text{d} V  &= G(n), \label{eq: diffusion integral constraint} \\
    \partial_r g (n,R) &= \frac{ k_{\text{poly}} n}{4 D}  g(n,R), \label{eq: diffusion R-BC}
\end{align}
\end{subequations}
with the diffusion constant, $D$, of actin monomers.
In Eq. \eqref{eq: diffusion R-BC}, a factor four remains from the surface area of the sink, while the factor $\pi R^2$ cancels out with the circular area of the branching zone when calculating the amount of consumed actin. 

This system can be solved analytically (see Appendix \ref{sec: solution local depletion}), yielding the monomer concentration at $r=R$
\begin{equation} \label{eq: local actin concentration exact}
    g(n) = \frac{G(n)}{V \left[ 1 + \frac{R k_{\text{poly}} n}{4 D}\left( 1 - \Delta \right) \right]},
\end{equation}
with a dimensionless geometrical factor, $\Delta$, which is defined in Eq. \eqref{eq: geometrical factor} and accounts for finite size effects.  The local depletion generates the factor $\sim 1/(1+n)$, which is of the nature proposed by Banerjee and Banerjee with an exponent of $-1$ in the size-feedback of the assembly rate \cite{banerjeeSizeRegulationMultiple2022}. A similar result for the growth of actin networks limited by diffusion was obtained by Paluch et al. \cite{paluchDeformationsActinComets2006}.

To derive an expression for the amount of remaining actin monomers in solution, $G(n)$, we need to know how many actin monomers are integrated into the network. To keep the number of variables as small as possible, we want to express the network length and its density profile solely in terms of the filament density $n$ in the branching zone. In order to do that, we again consider the steady state and make further simplifying assumptions. 

\subsection{Loss of percolation determines network length} \label{sec: Loss of percolation determines network length}
In general, structures subject to severing are not disassembled continuously, but instead lose larger pieces in a fragmentation process. While for single filaments this process can still be described in the framework of master equations \cite{rolandStochasticSeveringActin2008, rosarioUniversalLengthFluctuations2023, mominQuantitativeSignaturesDisassembly2025}, it is less clear how to approach this in the case of a network, where pieces break off when they become disconnected from the rest of the network. The loss of connectivity can be understood by percolation theory as a geometrical phase transition with the fraction of existing bonds as control parameter. If one removes bonds from a network, there is a critical fraction, $p_{\text{c}}$, at which the network is no longer percolated. Carlsson and Michalski proposed a lattice model as a coarse-grained representation of a branched actin network subject to severing \cite{carlssonDisassemblyActinNetworks2007, michalskiEffectsFilamentAging2010, michalskiModelActinComet2011}. 
Their numerical and analytical results agree very well with the observations made in reconstituted systems \cite{wiesnerBiomimeticMotilityAssay2003, manhartQuantitativeRegulationDynamic2019}, especially that the network breaks apart very rapidly, as soon as the connectivity falls below the percolation threshold. Therefore, we use the position at which the fraction of still intact actin filaments, $p_{\text{intact}}(l)$, drops below this critical value as an estimate for the length of the network, $L$, i.e. $p_{\text{intact}}(L) = p_{\text{c}}$. 
The critical value depends, in general, on dimension and topology of the considered network and is not precisely known for branched actin structures. However, it should be the same for actin networks with the same architecture regardless of their mesh size.

Since severing can occur at any position along a filament, the severing rate is proportional to its length \cite{mohapatraDesignPrinciplesLength2016}, which is, on average, given by the network's mesh size $\xi$. Thus, the total severing rate is $k_{\text{sev}} \xi$ with some effective base rate per unit length, $k_{\text{sev}}$, such that the fraction of intact filaments after time $t$ is given by $p_{\text{intact}}(t) = \exp(-k_{\text{sev}} \xi t)$.
When the system has reached its steady state, the network is growing with a constant velocity $v$ and we can convert time into distance to the branching zone, $l$, yielding 
\begin{equation} \label{eq: exponential decay}
p_{\text{intact}}(l) = \exp \left( -k_{\text{sev}} \xi l / v \right).
\end{equation}
The network length is then
\begin{equation} \label{eq: network length}
    L = \frac{v}{k_{\text{sev}} \xi } \ln \left( \frac{1}{p_{\text{c}}} \right),
\end{equation}
with the expected inverse proportionality to the mesh size $\xi$. This result was derived and verified with stochastic lattice simulations by Michalski and Carlsson \cite{michalskiModelActinComet2011}.

Like the branching rate $k_{\text{branch}}$ in Eq. \eqref{eq: filament density ODE}, the severing rate is an effective rate combining the rates of hydrolysis of ATP- to ADP-actin, the binding of ADF/cofilin and the subsequent filament severing. Thus, $k_{\text{sev}}$ implicitly contains the concentrations of ADF/cofilin and the recycling protein. One could try to derive estimates based on the protein concentration used in the experiment. However, the molecular details are not perfectly understood and rates have not been measured for some of these transitions. 
Additionally, the precise value of $p_{\text{c}}$ in Eq. \eqref{eq: network length} is not known. Therefore, we set $p_{\text{c}}$ to the value for a simple cubic lattice and will then choose these two effective rates, for branching and severing, as degrees of freedom in our model to match the experimental observables for a single network.

The elongation velocity of the network is obtained from the polymerization rate. In principle, the polymerization rate is force-dependent and can be described as a “Brownian ratchet” \cite{peskinCellularMotionsThermal1993}, typically used in the context of membrane protrusions \cite{mogilnerCellMotilityDriven1996, mogilnerRegulationActinDynamics2002}. However, the total mechanical load exerted by the bead is in the fN-regime (see Appendix \ref{sec: Relevance of the bead}), while the stall force for a single actin filament is a few pN \cite{peskinCellularMotionsThermal1993}. Therefore, we neglect the force-feedback such that $k_{\text{poly}}$ corresponds to the free polymerization rate \cite{pollardRateConstantsReactions1986}.
This also supports our assumption of a flat surface instead of the actual curved bead geometry because, without any significant load, the angle between filament and bead does not matter in the polymerization process. But the angle plays a role for the elongation speed of the network. We assume the perfect $\pm 35^\circ$ pattern because it is more likely under small load \cite{bielingForceFeedbackControls2016} and all filaments then contribute equally to the network elongation and stay in the branching zone. For the $0,\pm70^\circ$-pattern, the situation would be less clear. 
We introduce $\phi = \cos \left( 35^\circ \right) \approx 0.82$ as the projection onto the network direction to convert the polymerization velocity into the effective elongation velocity of the network
\begin{equation} \label{eq: polymerization velocity}
     v = \phi v_{\text{poly}} = \left\{\begin{array}{ll} 0, & n = 0 \\
         \phi k_{\text{poly}} g(n) d_0, & n > 0 \end{array}\right.
         \quad,
\end{equation}
with the half-size of actin monomers, $d_0$. We will see later on that the factor $d_0 \phi$ drops out and does not appear in our final equation \eqref{eq: nondim filament density ODE}. Therefore, the orientation does not matter for our main results. For completeness, the velocity is zero in the absence of a network.

Lastly, we compute the mesh size in Eq. \eqref{eq: network length} based on the filament density,
\begin{equation} \label{eq: filament spacing}
    \xi = \frac{1}{\sqrt{n}},
\end{equation}
such that $\xi$ corresponds to the average filament spacing. In Appendix \ref{sec: Mesh size of the network}
we discuss this simplifying assumption in more detail.

\subsection{Networks interact through a shared monomer pool}
\label{sec: Networks interact through a shared monomer pool}

In experiments, the actin density in comet tails reduces exponentially from the branching zone along its length \cite{wiesnerBiomimeticMotilityAssay2003} in agreement with the simple estimate in Eq. \eqref{eq: exponential decay}. However, the width of the network stays almost constant, i.e. no significant tapering takes place. Furthermore, in stochastic simulations without any depletion effects, the width has almost no impact on the network length if the network is initially significantly wider than its mesh size $\xi$ \cite{michalskiModelActinComet2011}. This allows us to estimate the total number of actin monomers in a network simply as the product of branching zone area $\pi R^2$ times the filament density integrated over the network length with a factor $1/(d_0 \phi)$ to convert the network length into number of monomers such that
\begin{align} \label{eq: F-actin in network}
\begin{aligned}
    F(n) &= \pi R^2 \frac{1}{d_0 \phi} \int_0^L n  \cdot p_{\text{intact}}(l) \text{d}l \\
    &= \pi R^2 \frac{n}{d_0 \phi} \int_0^L \exp \left(- \frac{k_{\text{sev}}}{\sqrt{n}} \frac{l}{v} \right) \text{d}l \\
    &= \frac{\pi R^2 k_{\text{poly}} \left( 1 - p_{\text{c}} \right)}{k_{\text{sev}}} g(n) n^{3/2},
\end{aligned}
\end{align}
where we have inserted the network length and velocity from Eqs. \eqref{eq: network length} and \eqref{eq: polymerization velocity} in the last step. The factor $1/(d_0 \phi)$ cancels out since the velocity depends on it as well. Note that the limit $p_{\text{c}} \to 0$, corresponding in Eq. \eqref{eq: F-actin in network} to an upper integration limit of 
$L \to \infty$, is well-defined. Thus, the assumption of a finite network length due 
to the loss of percolation is not crucial to obtain a finite value for the amount of actin in the network. However, we use a non-zero value for $p_{\text{c}}$ to obtain a finite network length as an additional quantity that can be compared to experimental data.
\textcolor{black}{Note that the 
exponential decay used in Eq. \eqref{eq: F-actin in network}
has been checked against experiments \cite{michalskiModelActinComet2011}
and hence this form implicitly includes the actin
that might have been polymerized at barbed ends that have been generated
by severing.}

For a single network, the number of remaining monomers in solution would be $G(n) = A_{\text{tot}} - F(n)$, where the initial number of actin monomers, $A_{\text{tot}}$, is known from the experimental setup \cite{guerinBalancingLimitedResources2025}. To generalize the approach to the case of multiple networks growing simultaneously, we assume that Eq. \eqref{eq: local actin concentration exact} still holds. Then the total amount of filamentous actin is simply the sum of the actin in each network. For $N$ networks, the remaining monomers are
\begin{equation} \label{eq: free monomers}
    G(n_1,...,n_N) = A_{\text{tot}} - \sum_{i=1}^N F_i(n_1,...,n_N), \\
\end{equation}
where the index $i$ represents the network of the $i$-th bead. By inserting Eqs. \eqref{eq: local actin concentration exact} and \eqref{eq: F-actin in network} this becomes a closed equation for $G(n_1,...,n_N)$ and can be solved explicitly
\begin{equation} \label{eq: free monomers final}
    G(n_1,...,n_N) = \frac{A_{\text{tot}}}{ 1 + \sum_{i=1}^N  \frac{ \pi R^2 k_{\text{poly}} \left( 1 - p_{\text{c}} \right) n_i^{3/2} }{V k_{\text{sev}} \left[ 1 + \frac{R k_{\text{poly}} n_i}{4 D}\left( 1 - \Delta \right) \right]}}.
\end{equation}

We have now derived all necessary quantities only in terms of the filament densities $n_i$ in the branching zone. Each network is described by an equation like Eq. \eqref{eq: filament density ODE} with the local monomer density given by Eq. \eqref{eq: local actin concentration exact} 
with the globally available monomers $G(n)$
replaced by Eq. \eqref{eq: free monomers final}.

\begin{figure}[t]
    \centering
    \includegraphics[width=0.45\textwidth]{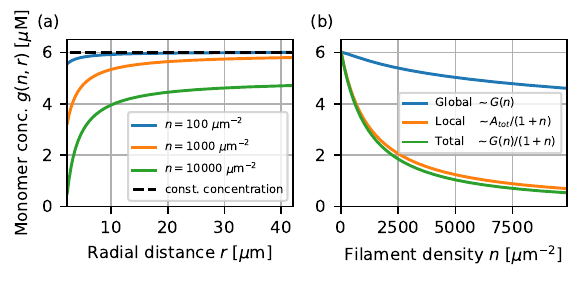}
    \caption{Effect of local monomer depletion. (a) Monomer concentration as a function of radial distance $r$ for different filament densities $n$, including local and global depletion. The dashed line corresponds to the constant concentration of $A_{\text{tot}} = 6 \ \mu$M in the absence of any monomer consumption. (b) Monomer concentration in the branching zone ($r=R$) as function of filament density $n$ considering only global depletion (blue curve), only local depletion (orange curve with $G(n) \equiv 6 \ \mu$M fixed), and the combination of both (green curve).}
    \label{fig: local depletion}
\end{figure}

To demonstrate the effect of local depletion, Fig. \ref{fig: local depletion}(a) shows the spatial concentration of actin monomers $g(n,r)$ for different filament densities in the case of a single bead. For a filament density in the order of 1000 $\mu \text{m}^{-2}$ (orange curve), typically found in actin networks \cite{abrahamActinbasedNanomachineLeading1999, plastinoEffectDiffusionDepolymerization2004, paluchDeformationsActinComets2006, kawskaHowActinNetwork2012, garnerLeadingEdgeMaintenance2022, colinRecyclingActinMonomer2023} and also later on in our model, the concentration in the branching zone is reduced by over 40$\%$ compared to far away from it. 
In turn, the monomer density has already reached 
over 90 $\%$ of its maximum value at a distance of 10 $\mu$m to the bead. In comparison, the microwell has a diameter of 100 $\mu$m, a height of 40 $\mu$m and the typical network length is roughly 20 $\mu$m, such that the average distance between beads and between the beads and the walls should be larger than the region around each bead where the depletion is significant. Thus, our approximations to obtain Eq. \eqref{eq: free monomers} are valid, as long as the number of beads is not too large. We verify this assumption by finite element method (FEM) simulations in section \ref{sec: Spatial model}.

In the limit $D \to \infty$, we obtain the expected result $g(n,R) \to G(n)/V$, such that the concentration would be the same everywhere. The scaling of the monomer concentration in the branching zone with $\sim 1/(1+n)$ can be seen in Fig. \ref{fig: local depletion}(b). By distinguishing between the global depletion of monomers via $G(n)$ and the local depletion caused by the factor $1/(1+n)$, we see that local depletion is the dominating effect, meaning that the growth of a network is diffusion-limited long before the global concentration would reach the critical concentration.

For further analysis, we non-dimensionalize these equations by rescaling the filament densities by $\left( V k_{\text{sev}} / \left( \pi R^2 k_{\text{poly}} \left( 1 - p_{\text{c}} \right) \right)\right)^{2/3}$ and time by $1/ \left( [\text{cap}] k_{\text{cap}} \right)$. Then, the dynamics of the dimensionless filament densities, $x_i$, are given by
\begin{equation} \label{eq: nondim filament density ODE}
    \dot x_i = \left( \frac{\mathcal{A}_i}{1 + \sum_{j=1}^N \frac{x^{3/2}_j}{1 + \mathcal{B}x_j}} \frac{1}{1 + \mathcal{B}x_i} - 1\right) x_i.
\end{equation}
This equation is the main result of our model derivation. It basically is our starting equation for growth,  Eq. \eqref{eq: filament density ODE},
but now closed by the known physics of growing actin networks. The first fraction within the parenthesis on the right-hand side
represents the global depletion of the finite pool corresponding to $G(n)$. 
It is more complicated than the usual linear expressions derived 
before for limited pool models \cite{mohapatraLimitingPoolMechanismFails2017} 
due to the physics of growing actin networks. The second fraction 
describes local depletion with the $\sim 1/\left(1+x\right)$ scaling discussed above. 
We consider Eq. \ref{eq: nondim filament density ODE} to be a general equation for the competition of growing
actin networks, because it represents the growth processes at play, but
does not require any specific molecular processes to lead to growth control and coexistence.
The problem is now described by a system of coupled ODEs and can be analyzed with tools from nonlinear dynamics. Because we have derived the central Eq. \eqref{eq: nondim filament density ODE} under steady state assumptions, in the following, we will focus on a bifurcation analysis to identify steady state solutions, i.e. the time derivative on the left-hand side will be set to zero.

Our theory has only two dimensionless parameters, which are defined as follows:
\begin{subequations}
\begin{align}
    \mathcal{A}_i &= \frac{A_{\text{tot}} [\text{NPF}]_i [\text{Arp}] k_{\text{branch}}} {V [\text{cap}] k_{\text{cap}}}, \label{eq: A_script} \\ 
    \mathcal{B} &= \left(\frac{k_{\text{poly}}} {R} \right)^{1/3} \left( \frac{V k_{\text{sev}}} { \pi \left( 1 - p_{\text{c}} \right)} \right)^{2/3} \frac{1 - \Delta}{4 D}. \label{eq: B_script}
\end{align}
\end{subequations}
The first parameter, $\mathcal{A}_i$, is an \textit{effective branching parameter} as it is given by the ratio between branching and capping and depends on the coating density $[\text{NPF}]_i$ of the corresponding bead. The second parameter, $\mathcal{B}$, is the same for all networks and combines the diffusion-limited assembly and the severing-based disassembly. Considering that $\mathcal{B}$ is inversely proportional to the monomer diffusion constant $D$ such that it weights the effect of local depletion in Eq. \eqref{eq: nondim filament density ODE}, it can be interpreted as an \textit{effective depletion parameter}. In particular, in the fast diffusion limit, $D \to \infty$, we obtain $\mathcal{B} \to 0$ and all local depletion terms vanish.

\begin{table*}[t]
\caption{\label{tab: Parameter values} Parameter values}
\begin{ruledtabular}
\begin{tabular}{llll}
 Symbol & Parameter & Value & Reference \\ 
\hline

 $A_{\text{tot}}$ & Total number of actin monomers & $1.134 \cdot 10^9$ & exp. setup in \cite{guerinBalancingLimitedResources2025} \\
 R & Bead radius & $2.25 \ \mu \text{m}$ & exp. setup in \cite{guerinBalancingLimitedResources2025} \\
 V & Microwell volume & $314 \ p \text{L}$ & exp. setup in \cite{guerinBalancingLimitedResources2025} \\
 $[\text{NPF}]$ & (high) NPF-coating density & 0.076 n$ \text{m}^{-2}$ & exp. setup in \cite{guerinBalancingLimitedResources2025} \\
 $[\text{Arp}]$ & Concentration of Arp2/3 complex & 90 nM & exp. setup in \cite{guerinBalancingLimitedResources2025} \\
 $[\text{cap}]$ & Concentration of capping proteins & 40 nM & exp. setup in \cite{guerinBalancingLimitedResources2025} \\
 $k_{\text{poly}}$ & Barbed-end monomer assembly rate & $11.6 \ \mu \text{M}^{-1} \text{s}^{-1}$ & measured in \cite{pollardRateConstantsReactions1986} \\
 $p_{\text{c}}$ & Percolation threshold of simple cubic lattice in 3D  & 0.2488 & taken from \cite{staufferScalingTheoryPercolation2021} \\
 $D$ & Diffusion constant of G-actin & $13 \ \mu \text{m}^2 \text{s}^{-1}$ & compare to \cite{raz-benaroushActinTurnoverLamellipodial2017, boujemaa-paterskiNetworkHeterogeneityRegulates2017} \\
 $d_0$ & Actin monomer half-size & 2.76 nm & taken from \cite{dominguezActinStructureFunction2011} \\
 $\phi$ & Network geometry factor & $\cos \left(35^\circ \right)\approx 0.82$ & compare to \cite{weichselTwoCompetingOrientation2010, mogilnerRegulationActinDynamics2002} \\
 $k_{\text{cap}}$ & Barbed end capping rate & 10 $\mu \text{M}^{-1} s^{-1}$ & measured in \cite{selveRateConstantCapping1986} \\
 $k_{\text{branch}}$ & Effective branching rate & $\approx 7.26 \cdot 10^{-11} \  \mu \text{m}^8 \text{s}^{-1}$ & comparison to exp. via Eq. \eqref{eq: A_script} \\
 $k_{\text{sev}}$ & Effective severing rate & $\approx 0.28 \ \mu \text{m}^{-1} \text{s}^{-1}$ & comparison to exp. via Eq. \eqref{eq: B_script} \\
 
\end{tabular}
\end{ruledtabular}
\end{table*}

All parameter values (see table \ref{tab: Parameter values}) are either given in the experimental setup \cite{guerinBalancingLimitedResources2025} or known from literature, except for the effective branching and severing rate. Because the parameter $\mathcal{A}$ only depends on $k_{\text{branch}}$ while $\mathcal{B}$ only on $k_{\text{sev}}$, we can determine them independently by choosing the two dimensionless parameters to match the experimental results. 

\section{Results} \label{sec: Results}

We now analyze our central equation Eq. \eqref{eq: nondim filament density ODE}
for the dimensionless filament densities $x_i$ of the $N$ competing networks.
Because of the autocatalytic nature of the branching process, there always exists the trivial solution, $x_i=0$, independently of the number and states of the other networks. The question is now under which conditions non-trivial solutions emerge 
and what they mean for the coexistence and
selection of dynamic actin networks.

\subsection{Single network and parameter estimates} \label{sec: Single network}

\begin{figure}[t]
    \centering
    \includegraphics[width=0.45\textwidth]{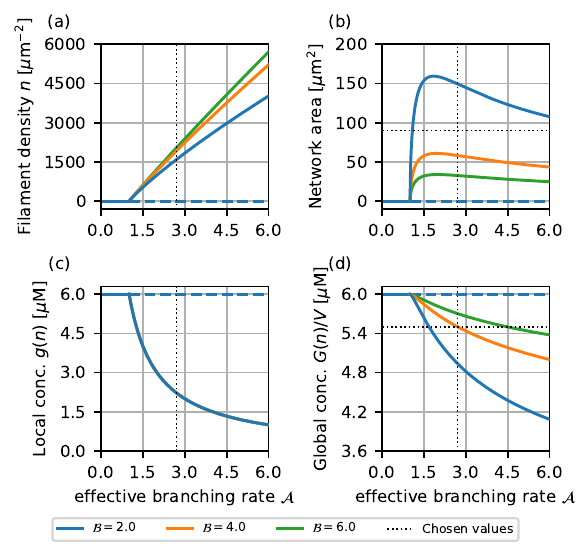}
    \caption{Bifurcation diagram for a single network obtained by numerical continuation. 
    (a) Filament density $n$ of the network in the branching zone as a function of
    effective branching rate $\mathcal{A}$ for different values of the 
    depletion parameter $\mathcal{B}$. The dimensionless parameters 
    $\mathcal{A}$ and $\mathcal{B}$ correspond to the rates $k_{\text{branch}}$ 
    and $k_{\text{sev}}$, respectively. 
    (b) Projected area of the network given by $2R$ times the network length $L$. 
    (c) Local concentration $g(n)$ of actin monomers in solution inside the branching zone, 
    independent of $\mathcal{B}$. 
    (d) Global actin monomer concentration as average over the whole microwell.}
    \label{fig: single bead bif}
\end{figure}

In the case of single network, the summation in Eq. \eqref{eq: nondim filament density ODE} disappears and the non-trivial steady states beyond $x=0$ are given by
\begin{equation} \label{eq: single network steady state}
    x^{3/2} + \mathcal{B}x + 1 - \mathcal{A} = 0.
\end{equation}
Since we only consider non-negative solutions, $x \geq 0$, and $\mathcal{B}$ is non-negative as well, the only possibility for a real  solution to occur is $1 - \mathcal{A} \leq 0$. Therefore, independent of $\mathcal{B}$, i.e. independent of the severing rate $k_{\text{sev}}$, we expect a non-trivial solution to exist for $\mathcal{A} \geq 1$, meaning that branching has to happen sufficiently fast compared to capping for the expression within the parenthesis in Eqs. \eqref{eq: filament density ODE} and \eqref{eq: nondim filament density ODE} to become positive and for a network to form. This is exactly what we find in the bifurcation diagram shown in Fig. \ref{fig: single bead bif}, obtained by numerical continuation \cite{doedelAUTO97Continuation1999}. To allow for direct comparison with experiments,
in the following the predictions of our model are always shown in dimensional units.
The filament density $n$ in Fig. \ref{fig: single bead bif}(a) monotonously increases for 
branching rate $\mathcal{A} \geq 1$. While the bifurcation point is independent of the depletion parameter $\mathcal{B}$, the filament density $n$ of the non-trivial solution increases with $\mathcal{B}$, corresponding to faster severing. This in turn leads to shorter networks incorporating less actin, according to Eq. \eqref{eq: F-actin in network}. Thus, there are more free monomers available corresponding to larger values of $G(n)$, as shown in Fig. \ref{fig: single bead bif}(d), enabling the growth of denser networks. In this sense, faster turnover allows for faster network growth by replenishment of the monomer pool.

Since the local concentration of actin monomers in the branching zone $g(n)$ is solely determined by the branching and capping process independently of the severing rate, we find that $g(n)$ in panel \ref{fig: single bead bif}(c) scales with $\mathcal{A}^{-1}$, i.e. with $k_{\text{branch}}^{-1}$, irrespective of the value of $\mathcal{B}$, exactly as expected.  Before and directly at the onset of network formation, i.e. directly at the bifurcation point, $G(n)$ and $g(n)$ are the same because the monomer concentration is homogeneous in the absence of any network acting as a sink.
The network area, given by $2RL$, in Fig. \ref{fig: single bead bif}(b) shows a non-monotonic behavior due to the competition of two effects. Faster branching leads to denser networks with smaller mesh sizes, which are, according to Eq. \eqref{eq: network length}, more stable against severing. Thus, the network becomes longer. However, denser networks also deplete the local actin pool (panel (c)) to a higher degree, reducing the polymerization velocity (Eq. \eqref{eq: polymerization velocity}). This in turn slows down filament and network elongation resulting in shorter networks.

To estimate the parameters $\mathcal{A}$ and
$\mathcal{B}$, we proceed as follows.
Because we are later on interested in the transition from coexistence to selection under increasing competition, $\mathcal{A}$ should be close to, but slightly larger, than the critical value. We also have to keep in mind that a “weak” bead with half the NPF-coating density (corresponding to a critical value of $\mathcal{A}=2$ at the same branching rate) should also be able to form a network in the absence of competition, i.e. $\mathcal{A}$ has to be larger than two.
We choose $\mathcal{A}=2.7$ and from Eq. \eqref{eq: A_script}, this corresponds to  $k_{\text{branch}} \approx 7.26 \cdot 10^{-11} \  \mu \text{m}^8 \text{s}^{-1}$. At this value, the filament density is between 1500 and 2000 $\mu \text{m}^{-2}$, corresponding to a filament spacing of roughly 22 to 26 nm, which agrees very well with values from literature \cite{abrahamActinbasedNanomachineLeading1999, plastinoEffectDiffusionDepolymerization2004, paluchDeformationsActinComets2006, kawskaHowActinNetwork2012, garnerLeadingEdgeMaintenance2022, colinRecyclingActinMonomer2023}. Based on Eq. \eqref{eq: polymerization velocity}, the polymerization and elongation velocity are directly proportional to the local monomer concentration $g(n)$. At the chosen value, we obtain $g(n) \approx 2.22 \ \mu$M and $v = \phi v_{\text{poly}} \approx \phi \cdot  4.26 \ \mu \text{m/min} \approx 3.50 \ \mu \text{m/min}$; also in very good agreement with the experimentally measured value of around $3 \ \mu \text{m/min}$ \cite{guerinBalancingLimitedResources2025}. We can also estimate the average filament length as the ratio of the filament elongation velocity and the capping rate,
\begin{equation} \label{eq: filament length}
    l_{\text{filament}} = \frac{v_{\text{poly}}}{[\text{cap}] k_{\text{cap}}} \approx \frac{4.26  \ \mu \text{m/min}}{0.4\  \text{s}^{-1}} \approx 177.5 \ \text{nm},
\end{equation}
which matches the values known from the literature \cite{bielingForceFeedbackControls2016, garnerLeadingEdgeMaintenance2022}. 

To constrain our second parameter $\mathcal{B}$, we can use two quantities provided by
Guérin et al. \cite{guerinBalancingLimitedResources2025}, namely the network length or area as structure size and the concentration of actin available in the bulk. The network area for a single bead per microwell is roughly $90 \ \mu \text{m}^2$ (see figure 2C in reference \cite{guerinBalancingLimitedResources2025}), while the remaining bulk monomer concentration, corresponding to $G(n)$ here, is around $5.5 \ \mu$M (see figure S1D in supplementary information of \cite{guerinBalancingLimitedResources2025}). We consider the monomer concentration as the more important quantity because it determines later on the competition strength between different networks. Thus, we set $\mathcal{B}=4.0$ to match the experimental value for $G(n)$ and obtain $k_{\text{sev}} \approx 0.28 \ \mu \text{m}^{-1} \text{s}^{-1}$ from Eq. \eqref{eq: B_script}. With this choice, the network length in our model is roughly 13 $\mu$m, the area accordingly $2RL \approx 60 \ \mu \text{m}^2$. We have approximated the actin network as perfect cylinders, while the actual network's shape is more complicated. Therefore, we overestimate the length by 
30 \%, while underestimating the area by about a third, but both values are in the correct order of magnitude
\cite{guerinBalancingLimitedResources2025}. 

Overall, with this choice of parameters, our model is able to reproduce all relevant values quantifying the state of the network. 
It should be kept in mind that the monomer concentration in the branching zone, $g(n)$, is only 40 $\%$ of the total concentration, $G(n)$, showing how significant local depletion is.

\subsection{Competition between identical networks} \label{sec: Competition between identical networks}

Having fixed all parameter values from our knowledge of single networks, 
we now analyze the competition between networks with identical assembly properties, i.e. for
the case of beads with the same NPF-coating density. In principle, each network is described by Eq. \eqref{eq: nondim filament density ODE} such that we have a system of $N$ equations for $N$ networks. As for the case of a single network, 
the non-trivial steady state
solutions are given by the expression within the parentheses. Since the effective branching parameter, $\mathcal{A}_i$, is the same for all networks, we can drop the index $i$. Then, the first fraction in Eq. \eqref{eq: nondim filament density ODE} is the same for all networks and basically represents the dimensionless form of $G(n_1,..,n_N)$, given in Eq. \eqref{eq: free monomers final}. Thus, we can write the steady state condition for each network as $\mathcal{A} / \left(1 + \sum_{j=1}^N \frac{x^{3/2}_j}{1 + \mathcal{B}x_j} \right) = 1 + \mathcal{B} x_i$.
Because the left-hand side is the same for all networks, we can equate the right-hand sides and directly obtain
\begin{equation} \label{eq: no symmetry broken states}
    x_i = x_j, \qquad \forall i,j \in \{1,...,N\}.
\end{equation}
Therefore, there are no symmetry broken states, where two or more networks form but of different densities. Obviously, one of the beads could not form a network at all. However, this state is unstable and in the only stable non-trivial state all the networks have equal density and size. This solution is well-defined and the degeneracy found in limited pool models without any size-dependent feedback \cite{mohapatraLimitingPoolMechanismFails2017} is lifted. In a stochastic simulation, one would expect only small fluctuations around this steady state as demonstrated by Banerjee and Banerjee \cite{banerjeeSizeRegulationMultiple2022}. We discuss the relevance of fluctuations in Appendix \ref{sec: Relevance of fluctuations}. 

We can now replace the sum in Eq. \eqref{eq: nondim filament density ODE} by a simple multiplication with the number of networks $N$ and all networks are described by the same density $x$ such that we can drop the index $i$ completely. The condition for a non-trivial steady state then reads
\begin{equation} \label{eq: mutiple same beads steady state}
    N x^{3/2} + \mathcal{B}x + 1 - \mathcal{A} = 0,
\end{equation}
where we can treat $N$ as a continuous parameter. 

Guérin et al. consider two different scenarios for equivalent beads in their experiments \cite{guerinBalancingLimitedResources2025}: “rapid” turnover corresponding to the conditions we used before, and “slow” turnover with half the concentration of both the disassembly protein ADF/cofilin and
the recycling protein. Both these proteins accelerate filament disassembly and we have absorbed them in the effective severing rate $k_{\text{sev}}$. For simplicity, we assume that $k_{\text{sev}}$ depends linearly on both of them such that, in the slow turnover case, $k_{\text{sev}}$ is reduced by a factor of 4 ignoring potential cooperative effects \cite{chanCofilinDissociatesArp22009}.

\begin{figure}[t]
	\centering
	\includegraphics[width=0.45\textwidth]{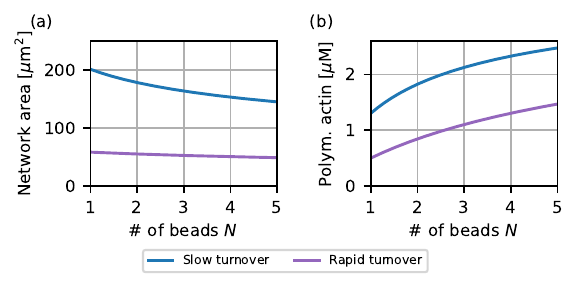}
    \caption{Competition between equivalent networks. (a) Projected network area per network and (b) total amount of polymerized actin within all networks as functions of the number of networks $N$. Rapid turnover corresponds to the value of the severing \textcolor{black}{rate} given in Table \ref{tab: Parameter values}. For slow turnover, the rate is four times smaller.}
    \label{fig: Competition between equivalent structures}
\end{figure}

In Fig. \ref{fig: Competition between equivalent structures}, we find that our model correctly reproduces the rapid turnover case where the network area is almost independent of the number of networks (Fig. \ref{fig: Competition between equivalent structures}(a)), while five networks polymerize roughly three times as much actin as one (Fig. \ref{fig: Competition between equivalent structures}(b)). Therefore, the networks experience only little competition. This changes in the slow turnover case, where the amount of polymerized actin does not even double from one to five networks, in good agreement with the experiment. It demonstrates the increased competition for slow turnover because each network contains more actin such that adding another bead has a larger impact on $G(n)$.
Even though we underestimate the area in general due to our parameter choice, its relative increase with a factor of around $3.5$ from rapid to slow turnover for a single network fits to the observed area increase. Again, our model underestimates the reduction in network area under stronger competition for slow turnover. Nevertheless, the general reduction of competition via turnover is captured well in our model, especially since we consider the amount of polymerized actin as more important.

\subsection{Coexistence of dense and sparse networks} \label{sec: Coexistence of dense and sparse networks}

Next we address the transition from coexistence of inequivalent structures to a state in which only the “strong” structures survive. In the bead motility assay, “weak” and “strong” structures can be created by using different NPF-coating densities on the beads, leading to slow and fast branching. Guérin et al. \cite{guerinBalancingLimitedResources2025} consider two species of beads, where the weak beads are coated with half the amount of NPFs. In our dimensionless formulation this corresponds to $\mathcal{A}_{\text{weak}} = 0.5 \mathcal{A}_{\text{strong}}$ and the critical monomer concentration for network formation is twice as large for the weak one.

\begin{figure}[t]
    \centering
    \includegraphics[width=0.45\textwidth]{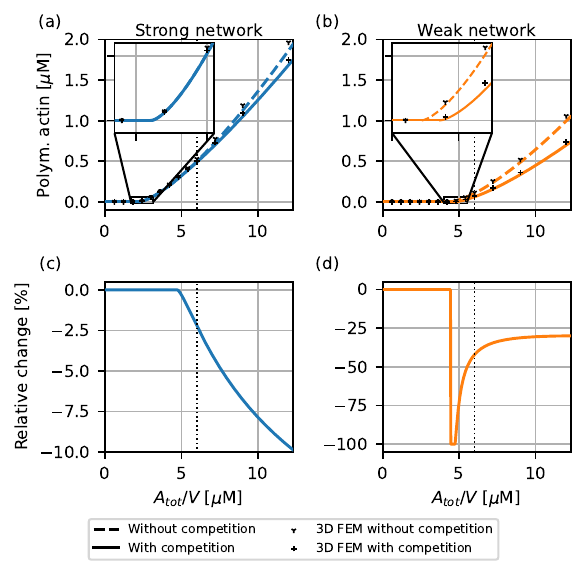}
    \caption{Coexistence of one weak and one strong network. (a) The amount of polymerized actin in the strong network without (dashed lines) and with competition (solid lines). The markers indicate FEM results of the spatial diffusion model.
    Insets show the respective bifurcation points associated with the onset of network formation. Dotted lines indicate the experimental value of $A_{\text{tot}}$ used in \cite{guerinBalancingLimitedResources2025}.
    (b) Same for the weak network. 
    (c) The relative change in network size between with and without competition
    for the strong network. (d) Same for the weak network. Note the different y-scale.}
    \label{fig: Two different beads bif diagramm}
\end{figure}

Fig. \ref{fig: Two different beads bif diagramm} compares the stable solutions for one weak and one strong bead in the same microwell (solid lines) to the situation in which they grow without competition (dashed lines). We have also included results obtained from a three-dimensional spatial model, which will be introduced and discussed in the next section \ref{sec: Spatial model}. We use the total amount of actin, $A_{\text{tot}}$, as the bifurcation parameter, because it is the most intuitive way to understand the onset of network formation and also is the easiest controllable parameter in the experimental setup. The strong bead (Fig. \ref{fig: Two different beads bif diagramm}(a)) is able to form a network at the same critical value of $g_{\text{strong}} = A_{\text{tot}}=2.22 \ \mu$M with and without competition, since the weak network has not started to grow yet. The amount of polymerized actin in the strong network is exactly the same in both cases, until the weak network (Fig. \ref{fig: Two different beads bif diagramm}(b)) starts to grow in the case of competition. We stress that in the presence of a strong network, the weak network is only able to grow due to the local depletion of actin monomers. Without, the bulk monomer concentration $G(n)$ would not further increase with $A_{\text{tot}}$, because all additional monomers would be integrated into the strong network. Therefore, the critical value $g_{\text{weak}}=4.44 \ \mu$M, at which the weak network could grow, would never be reached. This corresponds to the classical “winner takes it all” principle. With the assumption of local depletion, this critical value will be reached eventually. However, because a certain fraction of the monomers is bound within the strong network, the weak network starts to grow at a slightly larger value of $A_{\text{tot}}=4.72 \ \mu$M under competition, shifting the bifurcation point of the weak network to the right, highlighted in the inset of Fig. \ref{fig: Two different beads bif diagramm}(b). At this point, the curves of the strong network in panel \ref{fig: Two different beads bif diagramm}(a) start to differ such that the amount of polymerized actin is slightly less under competition because it also “feels” the presence of the weak network via the reduction of free monomers $G(n)$.

To quantify the effect of competition, the bottom row in Fig. \ref{fig: Two different beads bif diagramm} shows the relative difference in size for each network. Before the first bifurcation, at which the strong network starts to grow, the difference is zero, because there is no network formation in both cases. After the point, at which the weak network would grow without competition, the relative change for the weak network is maximum (100 \%). As soon as the weak network can also grow under competition, this difference reduces while at the same time the difference in strong network size starts to increase. Guérin et al. state a change in structure size of -60 \% for the weak one, while the strong network is not significantly affected by the competition \cite{guerinBalancingLimitedResources2025}. At the experimental conditions of $A_{\text{tot}} = 6 \ \mu$M, the relative change in polymerized actin in our model amounts to approximately -42.2 \% and -2.2 \% for the weak and strong network, respectively.

After having demonstrated the coexistence, we now investigate the behavior under an increase of competition strength. One way to achieve this is to add more strong beads. Based on our analysis in the previous section, we still only need two equations, one for the $N$ strong beads and one for the weak bead. One can actually reduce this further to a single equation by the same considerations made to obtain Eq. \eqref{eq: no symmetry broken states} and find that $x_{\text{weak}} = \left( x_{\text{strong}} + 1/\mathcal{B} \right) \mathcal{A}_{\text{weak}}/\mathcal{A}_{\text{strong}}  - 1/\mathcal{B}$ in the case of coexistence. However, to obtain the bifurcation points and the absolute values, we rely on numerical continuation. These results are again compared to those obtained by the spatial model, which will be introduced in section \ref{sec: Spatial model}.

\begin{figure}[t]
    \centering
    \includegraphics[width=0.45\textwidth]{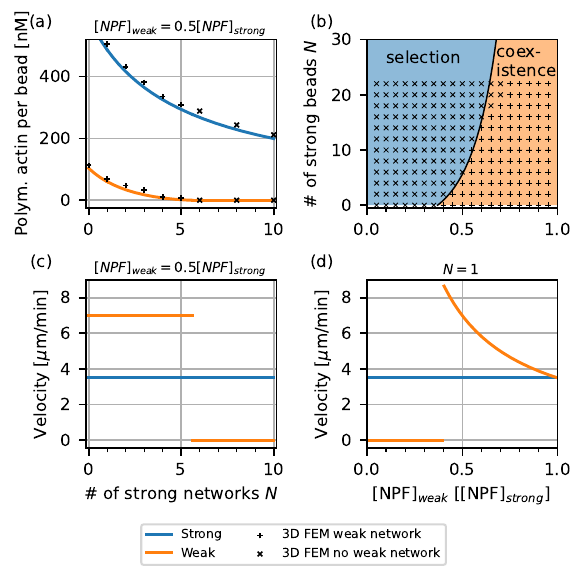}
    \caption{Transition from coexistence to selection for many networks. 
    (a) Amount of polymerized actin for one weak network in competition with $N$ strong ones. For $N \leq 5$, we observe stable growth of the weak network. For six or more, its growth is not possible anymore. 
    Markers indicate FEM results of the spatial diffusion model.
    (b) Phase diagram showing the two regions of coexistence and selection. The number of strong networks and ratio of NPF-densities are the two parameters defining the competition strength.
    (c) The network elongation velocity as function of the number of strong networks $N$.
    It is discontinuous at the bifurcation point.
    (d) The network elongation velocity as function of the weak NPF-coating density.}
    \label{fig: Transition from coexistence to selection}
\end{figure}

In Fig. \ref{fig: Transition from coexistence to selection}(a), we show how the number of strong networks present in the same microwell impacts the size of the weak network and its general ability to grow. As expected, the size decreases with increased competition until the weak network cannot grow anymore at all for six or more strong networks. This agrees very well with the experimental findings, where only a few percent of weak networks are able to grow for six strong beads and none for eight or more. In the experiment, Guérin et al. \cite{guerinBalancingLimitedResources2025}, placed several weak beads together with the $N$ strong ones in the same microwell and then counted the fraction of the weak beads which formed a network. This situation slightly differs from what we model, since we only consider one weak bead. However, the mechanism leading to the selection process is the same: more strong networks consume more actin monomers such that the number of free monomers $G(n)$ and thus also the local concentration at the weak bead reduce until it drops below the critical value of $g_{\text{weak}} = 4.44 \ \mu$M. 

The competition strength can also be controlled by the ratio of NPF-coating densities of the weak and strong beads. Even though variation in coating ratios has not been explored in the experiments, we can easily analyze this here. In Fig. \ref{fig: Transition from coexistence to selection}(b), the phase diagram for the growth of a weak network in the presence of strong networks is shown. The two considered parameters are the NPF-coating density of the weak bead and the number of strong networks. We find two regimes, stable growth of the weak network (coexistence) or no growth (selection). The maximum number of strong networks still allowing growth steeply increases with the NPF-coating density and diverges for $[\text{NPF}]_{\text{weak}} / [\text{NPF}]_{\text{strong}} \to 1$ because this corresponds to the case of equivalent networks. Even in the absence of any strong networks, $N=0$, below the critical value of $[\text{NPF}]_{\text{weak}} \approx 0.37 \cdot [\text{NPF}]_{\text{strong}} \approx 0.028 \ \text{nm}^{-2}$, the weak network cannot form because branching is too slow, i.e. the system is on the left side of the bifurcation in Fig. \ref{fig: single bead bif}. This point obviously depends on all parameters defining the branching parameter $\mathcal{A}$ in Eq. \eqref{eq: A_script}, especially on the amount of available actin $A_{\text{tot}}$.

So far, we have not discussed the behavior of the elongation velocity of the network. Based on Eq. \eqref{eq: polymerization velocity}, it is proportional to $g(n)$, but does not directly depend on the filament density $n$. Therefore, the moment a network is able to form, the velocity jumps to a finite value, i.e. it is discontinuous at the bifurcation point, as can be seen in Fig. \ref{fig: Transition from coexistence to selection}(c) and (d) for the weak network (orange curve).
In contrast, all other quantities like $G(n), \ g(n)$, and the network area are continuous (see Fig. \ref{fig: single bead bif}).
Since the steady state value of $g(n)$ of each network is solely determined by the balance of branching and capping, it does not depend on the number of networks $N$ in Fig. \ref{fig: Transition from coexistence to selection}(c), i.e. is independent of the competition strength. Thus, the velocity of the strong networks is constant. Likewise, the velocity of the weak network is constant before the bifurcation and twice as large. A general prediction of our model is that sparse networks are faster, because at a larger local actin concentration the rate of individual filament elongation and thus the elongation velocity of the whole network is larger.

When varying the NPF-density of the weak bead in Fig. \ref{fig: Transition from coexistence to selection}(d), we observe the same inverse relation of the velocity as before for the local actin concentration in Fig. \ref{fig: single bead bif}(c), since $v_{\text{weak}} \sim g_{\text{weak}} \sim 1/[\text{NPF}]$. Therefore, the velocity is the fastest right after the onset of network formation and subsequently decreases. For $[\text{NPF}]_{\text{weak}} = [\text{NPF}]_{\text{strong}}$, the two velocities are the same as we have reached 
the situation of two equivalent networks.

\subsection{Spatial diffusion model and dynamics}
\label{sec: Spatial model}

For our model derivation in section \ref{sec: Networks interact through a shared monomer pool}, we have calculated the steady state concentration profile of actin monomers for a single network and then made the simplifying assumption that it is still accurate in the case of several networks growing within the same microwell. However, for two or more networks, the problem no longer is spherically symmetric and cannot be solved analytically. We have also neglected the disassembling network as a source for monomers.

To demonstrate that these assumptions are valid and the predicted transitions still occur when including these effects, we now consider the fully time-dependent diffusion problem for $g$ coupled to the dynamic filament densities, $n_i$, and the dynamic amount of polymerized actin in each network, $F_i$. The dynamics of $n_i$ is still given by the ODE in Eq. \eqref{eq: filament density ODE}. $F_i$ increases by filament polymerization in the branching zone and decreases due to severing. This loss term then enters as a volumetric source in the diffusion equation of $g$. As described in section \ref{sec: Loss of percolation determines network length}, severing causes larger pieces to break off the network which then must be further depolymerized and recharged before being again available for polymerization. Because they will diffuse during this process, we simply assume a homogeneous source term across the whole microwell volume. However, a inhomogeneous source term would be straightforward to implement.  
Together with no-flux boundary conditions at the outer boundary of the microwell, this source term then ensures monomer conservation.
The monomer consumption by the networks is still modeled as spherical exclusions of radius $R$ within the domain subject to the same Robin boundary condition as before in Eq. \eqref{eq: diffusion R-BC}, 
\begin{equation} \label{eq: spatial R-BC}
    \mathbf{n} \cdot \nabla g = -\frac{ k_{\text{poly}} n_i}{4 D} g\ ,
\end{equation}
with the outward-pointing surface normal $\mathbf{n}$.
Because the monomer density might vary across these inner boundary surfaces, we define the average concentration across them as
\begin{equation} \label{eq: inner surface average}
    \hat{g}_i = \frac{1}{4 \pi R^2} \int_{\partial \text{Bead}_i} g(\mathbf{x}) \text{d}A\ .
\end{equation}
The coupled dynamic problem then reads
\begin{subequations} \label{eq: time-dependent spatial model}
\begin{align}
    \partial_t g(\mathbf{x},t) &= D \Delta g \ + \frac{1}{V} \sum_i^N \xi_i \tilde{k}_{\text{sev}} F_i, \label{eq: time-depndent diffusion} \\
    \partial_t n_i(t) &= \left ([\text{NPF}]_i [\text{Arp}] k_{\text{branch}} \hat{g}_i - [\text{cap}] k_{\text{cap}} \right) n_i, \label{eq: time-dependent filament density} \\
    \partial_t F_i(t) &= \pi R^2 k_{\text{poly}} \hat{g}_i n_i - \xi_i \tilde{k}_{\text{sev}} F_i. \label{eq: time-dependent F-actin}
\end{align}
\end{subequations}
Since $\hat{g}_i$ only appears in linear order in these equations, the averaging does not pose any issue. As before, the network mesh size, $\xi_i=1/\sqrt{n_i}$, is determined by the current filament density in the branching zone via Eq. \eqref{eq: filament spacing}. To account for the loss of percolation and derive the same steady state condition for $F$ as before, we define $\tilde{k}_{\text{sev}} = k_{\text{sev}} / ( 1 - p_{\text{c}})$. The weak formulation for the implementation \cite{wossnerCodeTheoryCoexistence2025} in 
the FEM software environment DUNE \cite{dednerGenericInterfaceParallel2010} can be found in Appendix \ref{sec: Numerical implementation of spatial model}.

\begin{figure}[t]
    \centering
    \includegraphics[width=0.45\textwidth]{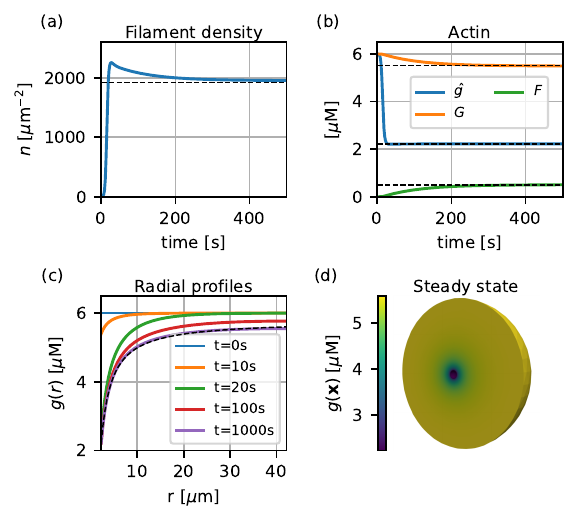}
    \caption{Full spatiotemporal treatment with FEM-simulations. 
    (a) Time evolution of the filament density as fast variable. The dashed black lines indicate the steady state obtained from the bifurcation analysis of the
    ODE-based theory.
    (b) Time evolution of actin quantities. $\hat{g}$ describes the average monomer concentration on the inner absorbing boundary. $G$ is the concentration of free monomers, $F$ the amount of polymerized actin within the network. Global monomer consumption ($G$) is slow compared to local consumption ($\hat{g}$).
    (c) Radial actin profiles at certain time points.
    (d) Steady state actin concentration shown for a cut through the center of the three-dimensional domain.
    }
    \label{fig: dynamics single bead}
\end{figure}

We first analyze the onset of network formation and determine the timescales on which branching, depletion and diffusion are occuring. Fig. \ref{fig: dynamics single bead} shows the time evolution for a single network placed in the center such that the geometry is the same as in our analytical solution. As initial conditions we chose a homogeneous G-actin concentration, $g(\mathbf{x},0) = A_{\text{tot}}/V$, such that $F=0 \ \mu$M, and a small but non-zero filament density $n=1 \ \mu\text{m}^{-2}$. The filament density in Fig. \ref{fig: dynamics single bead}(a) rapidly grows and, within 25 seconds, reaches its maximum by overshooting its steady state value slightly before relaxing towards the latter. On the same timescale, the local actin concentration $\hat{g}$ (blue curve in Fig. \ref{fig: dynamics single bead}(b)) gets depleted by the fast-growing network. As soon as $\hat{g}$ gets close to the critical concentration of $2.22 \ \mu$M, the branching process is slowed down such that the creation of new branches is balanced by capping, explaining the peak and subsequent relaxation of $n$. The local depletion of actin monomers can also be seen in Fig. \ref{fig: dynamics single bead}(c) showing the radial concentration profiles of actin, $g(r)$. The initial, flat profile decreases only close to the network (close to $r=0$) and reaches its minimum value there already at $t=20$ (coinciding with $\hat{g}$). The concentration far away from the network is not affected at this time point and is still the same as its initial value. Thus, the effect of monomer diffusion is significantly slower than the one of branching.

The amount of polymerized actin $F$ (green curve in Fig. \ref{fig: dynamics single bead}(b))
is strongly determined by filament density $n$ and starts to increase noticeably at $t \approx 15$ s, after $n$ has reached several hundred filaments per square micrometer. The subsequent slow growth of $F$ is accompanied by a slow global decrease of the actin monomers $G$, obeying monomer conservation $G + F = A_{\text{tot}}$. This slow process is also reflected in the evolution of the radial concentrations, whose tail far away from the network decreases towards the steady state profile in the order of ten minutes. Fig. \ref{fig: dynamics single bead}(d) shows a typical snapshot from the simulations close to this steady state.

From our FEM-simulations we can conclude that the system experiences a separation of timescales: fast branching and local depletion occurs within seconds, while network elongation and global consumption takes more than 10 minutes. The fast local adaptation justifies our approximation to neglect the movement of the bead: even at the maximum speed possible in our model of roughly 7 $\mu$m/min (see Fig. \ref{fig: Transition from coexistence to selection}(d)), the bead will never leave its self-created concentration hole. Furthermore, the steady state values obtained from the analytical result without the source term, i.e. from Eq. \eqref{eq: local actin concentration exact} as indicated by the dashed lines in Fig. \ref{fig: dynamics single bead}, are quite close to the three-dimensional simulation results. The FEM-approach with source term 
leads to a slightly (less than 1.5 \%) higher filament density.
This small difference can be understood by noting that the additional monomer source requires faster consumption by more filaments to achieve the same local concentration $\hat{g}$, which in turn leads to the same relative increase in the amount of polymerized actin $F$.

Having established the three-dimensional model and demonstrating excellent agreement with our previous result for a single network, we can now turn to the general case of multiple networks, already included in the Eqs. \eqref{eq: time-depndent diffusion}-\eqref{eq: time-dependent F-actin} via the index $i$. Only the spatial arrangement of the beads is open. For a good agreement with our approximation in section \ref{sec: Networks interact through a shared monomer pool}, the beads should have the maximum possible distance to each other but also to the outer boundary. For a few beads, this does not pose a problem. However, as already mentioned before, our approximation might break down for many beads. Besides the global depletion, for which we accounted for in Eq. \eqref{eq: free monomers final}, beads close to each other will also feel the local depletion. A weak network might not form in the vicinity of a strong one, even though the global actin concentration $G$ might actually be higher than its critical concentration. 
In general, the analytical approximation might overestimate the network densities but should not underestimate them.

\begin{figure}[t]
    \centering
    \includegraphics[width=0.45\textwidth]{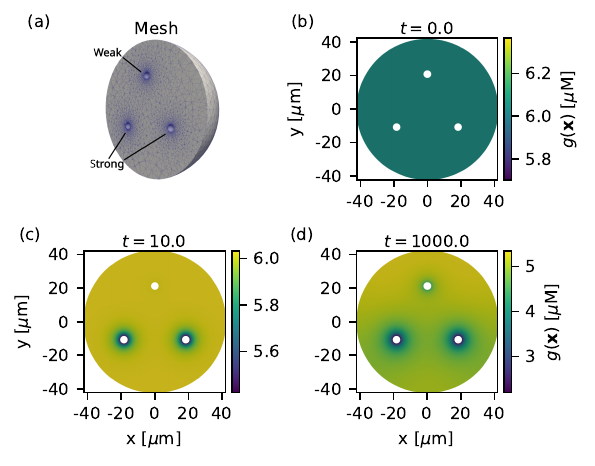}
    \caption{Spatial diffusion model for one weak network in competition with two strong ones. 
    (a) Cut through the computational mesh. The absorbing exclusion of the three networks are arranged in a triangle around the center of the sphere.
    (b)-(d) Actin monomer concentrations at different time points in the $z=0$ slice. The initial concentration in panel (b) at $t=0$ is homogeneous. In panel (d), the system has almost reached its steady state, in which all three networks coexist.
    }
    \label{fig: spatial 3 beads}
\end{figure}

Fig. \ref{fig: spatial 3 beads} exemplarily shows the case of three beads arranged in a triangle around the center of the spherical domain such that each bead has a distance of roughly 21 $\mu$m to both the center and the outer boundary. Fig. \ref{fig: spatial 3 beads}(a) shows the computational mesh, which is very fine at the inner absorbing boundaries to resolve the spherical exclusions and capture the steep gradients in their vicinity, while
becoming coarser further away from the exclusions. The upper bead is “weak”, while the two on the bottom are “strong”.
Fig. \ref{fig: spatial 3 beads}(b)-(d) show the temporal evolution of a two-dimensional slice cutting through the center of all three beads. We chose the same initial conditions as before with a homogeneous monomer concentration at $t=0$. After ten seconds, the strong networks have already grown and depleted their local environment to some degree. The concentration around the weak bead is almost the same as in the beginning. However, after 1000 seconds, representing basically the steady state, we can see local depletion around all three beads, i.e. also the weak network was able to form.

In Figs. \ref{fig: Two different beads bif diagramm} and \ref{fig: Transition from coexistence to selection},
one sees the excellent agreement between our theory and the results from the FEM-simulations.
For one strong and one weak network in competition in Fig. \ref{fig: Two different beads bif diagramm}(a) and (b), the loci of both bifurcations points, both with and without competition, agree perfectly between the two models, being best recognizable in the insets. Also, the amount of polymerized actin $F$ after the respective bifurcations shows very good agreement. As before, it is slightly larger in the spatial FEM model because of the source term. Overall, the approximations made for our central Eq. \eqref{eq: nondim filament density ODE} are negligible for the case of 
two networks which are sufficiently far away from each other.

To test this issue for more than two networks, we overlaid the FEM results in Fig. \ref{fig: Transition from coexistence to selection}(a), where the number of strong networks, which are in competition with one weak network, varies between zero and ten. While we treated it as a continuous parameter in Eq. \eqref{eq: mutiple same beads steady state}, the number of networks in the FEM simulations is actually discrete. For simplicity, we place the weak network in the center and arrange the strong networks around it. Again, the FEM simulations perfectly match the bifurcation point such that the weak network is still able to form for $N=5$ but not for $N=6$ anymore. Thus, even for in total seven networks, our approximation remains perfectly valid. 
The amount of polymerized actin depends on the specific spatial configuration of the networks, and hence deviates a little more or less for different network numbers $N$.

The FEM results even reproduce the transition threshold from selection to coexistence in Fig. \ref{fig: Transition from coexistence to selection}(b) for up to $N \leq 16$ almost perfectly. For more networks the transition shifts to the right, i.e. to higher NPF-densities of the weak bead, as expected. As explained before, the critical value varies to a small degree with the relative spatial positions of the networks. Placing the weak network further away from the strong networks or clustering the strong networks together, such that they impede each other's growth, would facilitate network formation at the weak bead, shifting the critical NPF-value to the left.
Nevertheless, the overall shape of the curve is the same in both models showing that, even for a very crowed environment, the simpler ODE-model makes qualitatively correct predictions and the found bifurcations occur in the same manner in the spatial model, which closely resembles the conditions in the experiment by Guérin et al. \cite{guerinBalancingLimitedResources2025}.

\textcolor{black}{
\subsection{Diffusion-limited assembly in the presence of an actin buffer}
\textit{In vivo}, the actin monomer concentration is actually several orders of magnitude higher than necessary for effective polymerization in reconstituted systems \cite{raz-benaroushActinTurnoverLamellipodial2017}. However, most of the actin monomers are held in a non-polymerizable state, because they are
sequestered by the actin monomer-binding protein thymosin \cite{cassimerisThymosinBeta41992, goldsteinThymosinV4Actinsequestering2005}. In principle, such a buffer might abolish the
limitation by diffusion, which is at the
heart of our theory.
To investigate the effects of such a buffer
in quantitative detail, we extend our FEM-model, given by Eqs. \eqref{eq: spatial R-BC} - \eqref{eq: time-dependent spatial model}, by including an additional concentration field for a buffer protein $b$, which can reversibly bind actin monomers in a 1:1 ratio, and another
concentration field for the actin-buffer complex $c$. The complex forms with rate $k_\text{ass}$ and dissociates with rate $k_\text{dis}$. This gives rise to an additional volumetric source and sink in the PDE for the actin monomer concentration. The three coupled PDEs are given by
\begin{subequations} \label{eq: spatial buffer PDEs}
\begin{align}
    \partial_t g(\mathbf{x},t) &= D \Delta g \ \ + \ k_\text{dis} c \ - \ k_\text{ass} g b \ + 
    \ \frac{1}{V} \sum_i^N \xi_i \tilde{k}_{\text{sev}} F_i, \\
    \partial_t b(\mathbf{x},t) &= D \Delta b  \ \ + \ k_\text{dis} c \ - \ k_\text{ass} g b, \\
    \partial_t c(\mathbf{x},t) &= \frac{D}{2} \Delta c \ - \ k_\text{dis} c \ + \ k_\text{ass} g b.
\end{align}
\end{subequations}
For simplicity, we assume the buffer protein to have similar size as monomeric actin, implying the same diffusion constant $D$, while the one of the complex is assumed to be only half as large. The boundary conditions for actin, $g$, are the same as before. To ensure a vanishing flux of the buffer and of the complex at the boundaries, we apply homogeneous Neumann conditions at the inner and outer boundaries. Note that due to the introduced reaction terms, the system of equations is now non-linear.}

\textcolor{black}{
The total amount of actin and buffer are determined by the initial conditions. In order to compare the situation for different amounts of buffer, $B_\text{tot}$, both with each other and to the case without buffer, we set the total amount of actin, $A_\text{tot}$, such that the equilibrium concentration of monomeric actin is exactly $G_\text{0}/V = 6 \ \mu$M in the beginning.
For a given equilibrium constant $K = k_\text{ass}/k_\text{dis}$, one can solve the steady state reaction equation to obtain
\begin{equation} \label{eq: total actin for given buffer}
    A_\text{tot} = \frac{K G_\text{0} B_\text{tot} + K G_\text{0}^2 + VG_\text{0}}{V + K G_\text{0}}.
\end{equation}
Then, the initial concentrations are $g_0 = 6 \ \mu$M, $b_0 = B_\text{tot}/V - (A_\text{tot}/V-g_0)$, and $c_0 = A_\text{tot}/V-g_0$.} \\

\textcolor{black}{
First, we analyze the situation of a single network (same situation as in Fig. \ref{fig: dynamics single bead}). How fast the reaction takes place, relative to diffusion, depends on the absolute values of the association and dissociation rate. For thymosin $\beta$4, these values are in the order of $k_\text{ass} \sim 1 ...10 \ \mu \text{M}^{-1} s^{-1}$ and $k_\text{dis} \sim 1 ...10 \ s^{-1}$ with an equilibrium constant $K \approx 1 \ \mu \text{M}^{-1}$ \cite{auWidelyDistributedResidues2008}. However, the buffer could be also realized by a different protein. Thus, we will later on vary the rates to cover a range of realistic values.} 

\begin{figure}[t]
    \centering
    \includegraphics[width=0.45\textwidth]{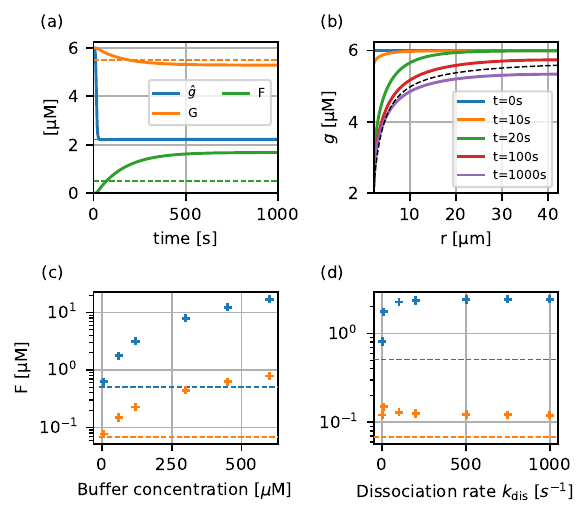}
    \caption{\textcolor{black}{Spatial diffusion model with a thymosin-like buffer. Total buffer concentration $B_\text{tot}/V = 60 \ \mu$M and reaction rates $k_\text{ass} = 10 \ \mu \text{M}^{-1} s^{-1}$ and $k_\text{dis} = 10 \ s^{-1}$. Dashed lines indicate steady state values without buffer.
    Single network:
    (a) Time evolution of actin quantities. 
    (b) Radial actin profiles at certain time points.
    One strong (blue) and one weak (orange) network in competition.
    (c) Varying the total buffer concentration for $k_\text{ass} = 10 \ \mu \text{M}^{-1} s^{-1}$ and $k_\text{dis} = 10 \ s^{-1}$.
    (d) Varying the reaction rates $k_\text{ass}$ and $k_\text{dis}$, while keeping the equilibrium constant $K$ as their ratio fixed, for $B_\text{tot} = 10 \ \mu$M.}
    }
    \label{fig: buffer model}
\end{figure}

\textcolor{black}{
Fig. \ref{fig: buffer model}(a) exemplarily shows the time evolution for $k_\text{ass} = 10 \ \mu \text{M}^{-1} s^{-1}$, $k_\text{dis} = 10 \ s^{-1}$, and a total buffer concentration of $B_\text{tot}/V = 60 \ \mu$M. As before for the case without buffer, the free actin monomer concentration at the bead surface, $\hat{g}$, decreases as the network grows. Then the actin-buffer complexes start to dissociate and replenish the free actin available for polymerization. Because of this additional actin getting released, roughly three times as much actin is incorporated into the network compared to the situation without buffer, when the system has reached its steady state. Thus, the presence of a buffer actually allows the network to grow to a larger size and is able to partially compensate for the diffusion limit. As a consequence, the total amount of free actin $G$ in steady state is smaller with a buffer present. However, the system still experiences spatial gradients, which are only slightly less steep, and actin is still strongly depleted at the bead locally, see Fig. \ref{fig: buffer model}(b). The concentration of the complex $c$ has the same spatial profile, while the buffer concentration is inverted and the highest at the bead. Thus, the buffer and complex concentrations are limited by diffusion themselves and cannot prevent spatial inhomogeneities.} 

\textcolor{black}{
To demonstrate that the presence of a buffer does not change the main conclusions of our theory, we analyze the same setup as in Fig. \ref{fig: Two different beads bif diagramm}, with one strong and one weak network in competition. We increase the buffer concentration up to 600 $\mu$M and the total actin concentration, according to Eq. \eqref{eq: total actin for given buffer}, up to $A_\text{tot} = 520 \ \mu$M, which is presumably the order of magnitude present in cells \cite{raz-benaroushActinTurnoverLamellipodial2017}. In Fig. \ref{fig: buffer model}(c), the amount of F-actin in the strong network (blue) grows roughly linearly with the buffer concentration (note the logarithmic scale). The weak network (orange) profits from the buffer as well, but less strongly. 
}

\textcolor{black}{
When speeding up the actin-buffer reaction while keeping the buffer concentration fixed, cf. Fig. \ref{fig: buffer model}(d), 
the strong network size initially grows, but then plateaus and does not increase further. Even when the reaction is slow ($k_\text{ass} = 0.1 \ \mu \text{M}^{-1} s^{-1}$ $k_\text{dis} = 0.1 \ s^{-1}$) compared to actin polymerization ($k_\text{on} = 11.6 \ \mu \text{M}^{-1} s^{-1}$), the filament density is twice as high as without buffer. However, speeding up the reaction only leads to a further subsequent growth of 25 \%, even at reaction rates 10.000-fold faster. At the rates used before in the panels (a)-(c), the strong network size has already reached more than 90 \% of its maximum value. Thus, the system is still diffusion-limited, since the actin-buffer complex is locally depleted in the same manner as the actin monomers. In contrast, the weak network size scales non-monotonically with the reaction speed and is the largest at intermediate reaction rates ($k_\text{dis} \approx 1.0$). Nevertheless, for all combinations of buffer concentration and reaction speed, the weak network is still able to form and is larger than without buffer. Thus, within our theory, coexistence of fast and slowly branching networks is still possible with a buffer, regardless of its concentration or molecular details. The position of the bifurcation point, at which the switch from selection to coexistence happens (compare to onset of weak network formation in Fig. \eqref{fig: Two different beads bif diagramm}(b)), is almost unaffected by the buffer.
However, for high buffer concentrations, the filaments would become so densely packed that, at some point, other factors would become size limiting. Either other proteins involved in branching, like NPF or Arp2/3, are getting depleted or the filaments would impede each other due to steric interactions between them.
}

\section{Discussion} \label{sec: Discussion}

Here, we have derived a general theory for branched actin networks growing simultaneously
in a shared and limited pool of actin monomers. Our theory shows that monomer depletion in the growth region is the relevant feedback mechanism that allows for coexistence of the different networks. 
\textcolor{black}{Our work provides a detailed realization of a more abstract theory for the effect of limited pools previously proposed \cite{banerjeeSizeRegulationMultiple2022}
and explains central aspects of recent experimental observations
by Guerin et al. \cite{guerinBalancingLimitedResources2025}.
For a single network, we obtain values for the filament density, filament length, network length, and bead velocity in the correct orders of magnitude or even close to the actual measured values.
In the slow turnover case, we get a correct relative increase in network area and incorporated actin for a single network.
For the competition of one weak and one strong bead, we obtain a relative decrease of 42 \% for the weak network, that is in a similar range to the 60 \% reported by Guerin et al. Finally, the transition from coexistence to selection at $N=6$ is the one observed experimentally.} 

In more detail, for a single network our model results in a bifurcation to growth for sufficiently fast branching. 
By comparing the amount of actin polymerized into the network to the experimental measured values \cite{guerinBalancingLimitedResources2025}, \textcolor{black}{we could fix the two effective parameters of the theory, without
any need for detailed fitting.} For these parameter values, the concentration of actin monomers in the branching zone is significantly lower than far away and local depletion is the main limiting factor for growth instead of global depletion of monomers. Thus, even if other potential feedback mechanisms, which are not considered by us, are affecting the growth properties, these findings demonstrate that local depletion is significant and has to be taken into account for fast-growing structures.

In the case of several equivalent networks, we find well-defined steady states, different from corresponding theories for linear
actin structures \cite{mohapatraLimitingPoolMechanismFails2017}. We also see the experimentally observed sublinear scaling of the total amount of polymerized F-actin with the number of networks, showing the effect of competition for the same pool, and the reduction of competition by increased turnover. 
\textcolor{black}{
However, our model underestimates the relative changes in network area with the number of competing networks, especially for slow turnover. As we assume a cylindrical tail shape, the network area depends linearly on its length, while the true shape observed in the experimental microscopy images is more complex. 
Because competition is determined by the available actin monomers, we consider the amount of polymerized actin as the most important observable. The values of the free parameters in our model have been fixed to reproduce the scaling of F-actin with the filament density, which is the central quantity in our theory. By assuming an idealized tail shape, a simple exponential density profile along the tail and determining the network length by a single percolation criterion, the dependence of the network area on the filament density and its change under variation of turnover are captured only qualitatively by these approximations. This is also reflected by the fact that
the geometrical factor $\phi$ drops out of our main equation.
In future work, one could improve on this part of the 
theory by including more details on the actin network geometries.}

For two beads with different NPF-coating densities, we find the possibility of coexistence, and our model yields correct values for the relative changes in structure size with and without competition. The fast branching (“strong”) network is almost not affected at all, while the slow (“weak”) network is only able to grow to roughly half its former size.
Strikingly, our theory also explains the switch from a regime of coexistence to selection under increasing competition, with the transition occurring at the same number of networks as found experimentally \cite{guerinBalancingLimitedResources2025}. We predict how changing experimentally controllable parameters would shift the critical number of networks and present the full phase diagram for coexistence and selection.
Furthermore, our hypothesis that the system is mainly diffusion-limited by local actin depletion leads to three predictions for the steady state bead velocities, which allow for experimental verification: first, the velocity is discontinuous at the bifurcation or at least increases very steeply with the control parameter. Second, sparser networks are elongating faster because they cause less local depletion. Thus, the bead velocity is inversely related to the NPF-coating density of the bead. 
Because of this, in our model, sparse networks can be longer despite being more vulnerable to severing. 

Lastly, the velocity depends only on the individual assembly properties, like the NPF-coating density, but not on global quantities, like the total number of available actin monomers. Therefore, the bead velocities are independent of the competition strength, defined by the number of competing networks or the turnover rates.
In principle, it would also be possible to invert the relation and estimate the local monomer concentration based on the bead velocity, which is easier to measure.
\textcolor{black}{Guerin et al. \cite{guerinBalancingLimitedResources2025} did not report individual velocities for beads with different coating densities, which would allow for a direct comparison to our predictions.
However, direct experimental support for a negative correlation of NPF density and growth speed, due to local actin monomer depletion, comes from the growth of “lamellipodium-like structures” from 2D substrates \cite{boujemaa-paterskiNetworkHeterogeneityRegulates2017}. There, growth is slowed down under increasing NPF density of nucleation spots on bar shaped patterns of constant geometry. Furthermore, actin fluorescence intensity (a proxy for network density) was shown to be negatively correlated with velocity, while network geometry and mechanical load remain unchanged. These findings are in good agreement with our model. Because these experiments were performed on 2D substrates, we note that monomer diffusion is expected to be more hindered than in 3D bead assays, potentially amplifying the diffusion-limiting effect.}

\textcolor{black}{
Here we have assumed that only the local actin monomer concentration determines the bead velocity. In practice, other factors might also contribute \cite{mogilnerEdgeModelingProtrusion2006}. Our continuum description could also break down for very sparse networks, where only a few filaments are propelling the bead and longer filaments may easily bent due to their reduced stiffness.
Nevertheless, we expect that the predicted reduction in velocity with increasing NPF coating density, arising from diffusion-limited monomer supply, constitutes an effect that should be superimposed on the velocity curve and remain observable in corresponding measurements. In general, it would be highly interesting
to systematically study the relation between network density and
bead velocity in future experiments.
}

One of the main approximations in our model is to use the monomer concentration profile, derived for a single network, also in the case of competition. To demonstrate the validity of this assumption, we have presented an extension of our model in which we numerically solve the diffusion equation in three dimensions to obtain the monomer concentration for any number and spatial arrangement of beads. We find excellent quantitative agreement with the results from the ODE-based theory for the experimentally studied number of beads and microwell volume, and only small deviations for more. Thus, our central equation works very well in the regime investigated in the bead motility assay \cite{guerinBalancingLimitedResources2025} of at most ten beads.
The distances between the beads, especially between the weak and strong beads, has a large impact on the ability of the weak network to grow. For our simulations, we chose an arrangement for which the beads were more or less uniformly spaced with several micrometers distance to the outer boundary. Beads which are very close together, impair each other's growth. A strong bead in the vicinity of a weak one can deplete the local monomers to such a degree that the weak network cannot form. Thus, the relative positions present another degree of freedom, smearing out the sharp transition in the simplified model.
This might explain why in the bead motility assay \cite{guerinBalancingLimitedResources2025} some weak beads are able to form a network while others, with the same NPF-coating density subject to the same competition strength, are not.

The spatial model also enables us to analyze dynamic behavior. We find a separation of timescales between different processes: branching and local depletion occur in the orders of a few tens of seconds, while network elongation and global depletion takes several minutes. The bead velocity of $3.5-7 \ \mu \text{m/min}$ is slow compared to local depletion such that neglecting the movement of the bead in our model is justified. In the derivation of our model, we have made further assumptions, which might lead to a discrepancy to measured values. Especially right after the bifurcation, our continuum approach might fail for very sparse networks. Assumptions regarding the network formation and branching process are discussed in more detail in Appendix \ref{sec: Model assumptions}.

\textcolor{black}{
When including a reversible actin–buffer system into our analysis, we found that a large reservoir of sequestered monomers can partially alleviate diffusion-limited growth by locally releasing polymerizable actin. This increases overall network size and filament densities, particularly for fast-growing networks. However, even for high buffer concentrations and rapid reaction kinetics, spatial gradients in both free actin and actin–buffer complexes persist, indicating that the system remains fundamentally diffusion-limited. Consequently, the coexistence of strong and weak networks 
is a fundamental property that is robust against the presence of a buffer.}

\textcolor{black}{
At very high buffer and actin concentrations, additional mechanisms are likely to become limiting. In particular, depletion of other essential components (e.g., NPFs or Arp2/3), steric hindrance between densely packed filaments 
and reduced actin monomer diffusion in a porous environment
may dominate over local monomer depletion. In this regime, coexistence may arise from these alternative constraints rather than diffusion-limited supply of actin alone.}

\textcolor{black}{
Our assumption that all non-network actin is immediately polymerizable 
is challenged by the possibility that some of it might persist
in short actin filaments. This situation can be modeled
by introducing a pool of oligomeric actin \cite{raz-benaroushActinTurnoverLamellipodial2017}. 
By subtracting the amount of oligomers on the right-hand side of Eq. \eqref{eq: free monomers}, one gets a reduction of 
the available monomer concentration and enhancement of the competition between networks, leading to lower steady-state filament densities. The size of such a non-polymerizable actin pool would be controlled by the ratio of the severing rate in our model and an effective depolymerization rate, characterizing the timescale on which oligomers are broken down into individual monomers.
Nevertheless, when assuming simple kinetics for the oligomer turnover,
the qualitative structure of the governing equations and the existence of the coexistence–selection transition are not altered.
For a single network, the steady state equation is of the same form as Eq. \eqref{eq: single network steady state} with a constant 
depending on the depolymerization rate being added to $\mathcal{B}$.
Thus, while quantitative details such as filament density and network size may shift, the central mechanism remains intact.}

Even though we have developed this model to specifically explain the observations in the recent publication by Guérin et al. \cite{guerinBalancingLimitedResources2025}, the role of protein depletion in intracellular assembly as regulatory mechanism is much more general. It would be interesting to investigate, both experimentally and theoretically, how local depletion of actin effects the assembly of different architectures, like actin cables or stress fibers for which negative feedback between cross-sectional area and elongation speed should exist. In combination with the depletion of other actin binding proteins, like Arp2/3 or ADF/cofilin, this might play an important role in the observed coexistence of the various different actin architectures in cells and might eliminate the need for additional size-sensing mechanisms.

\section*{Acknowledgments} 

We thank Peter Bastian for helpful discussions on the FEM software environment DUNE. This work is funded by Deutsche Forschungsgemeinschaft (DFG, German Research Foundation) under Germany’s Excellence Strategy - EXC 2181/1 – 390900948 (the Heidelberg STRUCTURES Excellence Cluster).

\appendix

\textcolor{black}{
\section{Spatially resolved model of actin network} \label{sec: Spatial model of actin network}
To better motivate Eq. \eqref{eq: filament density ODE}, we here formulate a one-dimensional spatial model along the actin tail including both capped and uncapped filaments, $n_c$ and $n_u$.
As explained in section \ref{sec: Molecular mechanisms underlying network growth}, NPFs are surface bound and activated Arp2/3 is only available close to the bead surface. We thus consider the branching rate to decay exponentially with the distance $x$ from the bead surface with a characteristic length $l_0$, representing the extension of the branching zone.
We account for local depletion due to diffusion-limited actin monomer supply by a local factor $A_\text{tot}/(1 + R k_\text{poly}n/4/D)$ (global depletion is not relevant for the discussion here). Similar to Eq. \eqref{eq: filament density ODE}, the dynamics of uncapped filaments is given by
\begin{equation} \label{eq: uncapped filaments spatial}
    \dot{n}_u = \tilde{k}_\text{branch} e^{-x/l_0} \frac{n_u + n_c}{1 + \frac{R k_\text{poly}}{4 D} n_u} - \tilde{k}_\text{cap} n_u,
\end{equation}
but new branches can also be created from already capped filaments. For the sake of clarity, we have absorbed the protein concentrations into the respective rates, $\tilde{k}_\text{branch} = {k}_\text{branch} [\text{NPF}] [\text{Arp}] A_\text{tot}$ and $\tilde{k}_\text{cap} = {k}_\text{cap} [\text{cap}]$. 
Conversion of ATP-actin to ADP-actin, binding of ADF/cofilin and subsequent severing takes time,
thus severing does not take place
in the branching zone and can be neglected here.
Eq. \eqref{eq: uncapped filaments spatial} does not contain any derivatives and is only quadratic in $n_u$. It thus can be solved analytically in dependence of the capped filament density $n_c$. Because the latter do not elongate and are pushed backwards by the polymerizing uncapped filaments, i.e. they are advected, their dynamics is given by
\begin{equation} \label{eq: capped filament PDE}
    \dot{n}_c = \tilde{k}_\text{cap} n_u - v_0 \partial_x n_c.
\end{equation}
The advection velocity is given by Eq. \eqref{eq: polymerization velocity} for $n_u$ at the left boundary, $v_0 = v(n_u(0))$. 
Filaments are immediately pushed backwards after being capped (as long as $v_0 \neq 0$). Therefore, we close this equation by a homogeneous Dirichlet condition at the left boundary, $n_c(x=0) = 0$. This boundary value problem has to be solved numerically, since the solution for $n_u$ depends nonlinearly on $n_c$. However, 
to derive the steady state solution, formally we can integrate Eq. \eqref{eq: capped filament PDE} and obtain
\begin{equation}
    n_c(x) = \frac{\tilde{k}_\text{cap}}{v_0} \int_0^x n_u(x,n_c) \ \text{d}x.
\end{equation}
Thus $v_0$ scales the ratio of capped and uncapped filaments. 
}

\begin{figure}[t]
    \centering
    \includegraphics[width=0.45\textwidth]{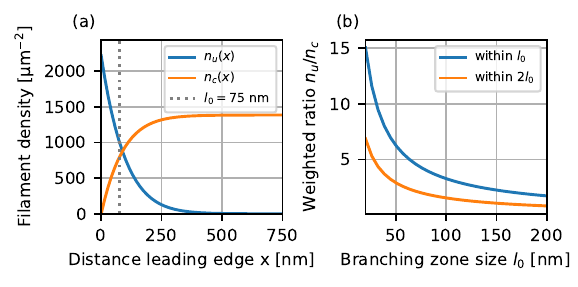}
    \caption{\textcolor{black}{Spatial model of actin network.
    (a) Steady state solutions of Eqs. \eqref{eq: uncapped filaments spatial} and \eqref{eq: capped filament PDE} for $l_0 = 75$ nm.
    (b) Exponentially weighted ratio of uncapped and capped filaments $\int u_u e^{-x/l_0} \text{d} x / \int u_c e^{-x/l_0} \text{d} x$ as relative contribution to branching for $x \in [0,l_0]$ (blue curve) and $x \in [0,2l_0]$ (orange curve).
    }}
    \label{fig: spatial actin tail model}
\end{figure}

\textcolor{black}{
A numerical solution for the characteristic length $l_0 = 75$ nm is shown in Fig. \ref{fig: spatial actin tail model}(a). The uncapped filament density is the highest at $x=0$ and then decreases because of the exponentially decaying branching rate. The capped filament density is zero at the beginning due to the boundary condition and then increases because more and more filaments are getting capped over time.
To quantify the contributions of capped and uncapped filaments to the branching process in the relevant zone, we integrate both densities in the interval $x \in [0,l_0]$ weighted by the exponential factor $e^{-x/l_0}$. The weighted ratio is around $4.3$, meaning that for the parameter values used in our ODE model (Table 1), the uncapped filaments are actually dominating and contributing more than four times as much to branching than the capped filaments.
The blue curve in Fig. \ref{fig: spatial actin tail model}(b) shows how this ratio changes with the length $l_0$ of the branching zone. As expected, the smaller $l_0$ is, the less relevant are the capped filaments and the ratio diverges for $l_0 \to 0$. Even if one considers twice the extension of the branching zone, $2 l_0$, the uncapped filaments are still dominating. }

\textcolor{black}{
For different sets of parameter values, this is not necessarily the case and the solutions can change even qualitatively. For example, if the branching rate $\tilde{k}_\text{branch}$ would be larger, the advection velocity $v_0$ would be smaller and, counterintuitively, the capped filaments would become more relevant and might dominate.
Nevertheless, the approximation to only consider uncapped filaments in Eq. \eqref{eq: filament density ODE} is self-consistent, because for the value of $k_\text{branch}$ obtained via Eq. \eqref{eq: A_script} in section \ref{sec: Single network}, the uncapped filaments are actually dominating in the here presented spatial model. 
In general, the ODE Eq. \eqref{eq: filament density ODE} can be considered as a limit of the one-dimensional PDE \eqref{eq: uncapped filaments spatial}, either for a thin branching zone, $l_0 \to 0$, or for a fast network elongation, $v_0 \to \infty$.
}

\section{Local depletion with monomer conservation} \label{sec: solution local depletion}
We here solve the one-dimensional boundary value problem for the monomer concentration, given by Eqs. \eqref{eq: diffusion PDE SS}-\eqref{eq: diffusion R-BC}. The Laplace Eq. \eqref{eq: diffusion PDE SS} can be solved directly by integrating twice, leading to two integration constants, $a$ and $b$, in $g(n,r) = -a/r + b$. These constants are uniquely defined by the integral constraint \eqref{eq: diffusion integral constraint} and the Robin boundary condition \eqref{eq: diffusion R-BC}.
Plugging the general solution into the integral and transforming to spherical coordinates yields
\begin{align}
    \begin{aligned}
       \frac{G(n)}{4 \pi} &= \int_R^Z  \left(-\frac{a}{r} + b \right) r^2 \text{d} r \\
       &= -\frac{a}{2} \left( Z^2 - R^2 \right) + \frac{b}{3} \left( Z^3 - R^3 \right).
    \end{aligned}
\end{align}
The outer radius is determined by the microwell volume in the experiments (see table \ref{tab: Parameter values}) via  $V = 4 \pi/3 \ (Z^3 - R^3)$ and is $Z = 42.17 \ \mu$m.
We express $b$ in terms of $a$
\begin{align} \label{eq: prefactor b diffusion equation monomer conservation}
\begin{aligned}
    b &= \frac{3}{Z^3 - R^3} \left[ \frac{G(n)}{4 \pi} + \frac{a}{2} \left( Z^2 - R^2 \right)  \right] \\
    &= \frac{G(n)}{V} + \frac{3a}{2} \frac{Z^2 - R^2}{Z^3 - R^3} \\
    &= \frac{G(n)}{V} + \frac{a \Delta}{R},
\end{aligned}
\end{align}
where we defined the geometrical factor $\Delta$ in the last line,
\begin{equation} \label{eq: geometrical factor}
    \Delta = \frac{3R}{2}\frac{Z^2 - R^2}{Z^3 - R^3} \approx 0.08.
\end{equation}
This expression becomes zero in the limit of a large volume corresponding to $Z \to \infty$.
We then evaluate the boundary condition \eqref{eq: diffusion R-BC}
\begin{align}
    \begin{aligned}
        \partial_r g(n,R) &= \frac{a}{R^2} \\
        & {\overset{!}{=}} \frac{ k_{\text{poly}} n}{4 D}  g(n,R) = \frac{ k_{\text{poly}} n}{4 D} \left[ -\frac{a}{R} + b \right] \\
        &= \frac{ k_{\text{poly}} n}{4 D} \left[ -\frac{a}{R} + \frac{G(n)}{V} + \frac{a \Delta}{R} \right],
    \end{aligned}
\end{align}
yielding a closed expression for $a$
\begin{equation} \label{eq: prefactor a diffusion equation monomer conservation}
    a = \frac{G(n) R^2}{V} \left[ \frac{4 D}{ k_{\text{poly}} n} + R \left( 1 - \Delta \right) \right]^{-1}.
\end{equation}
The solution is fully determined by the Eqs. \eqref{eq: prefactor b diffusion equation monomer conservation} and \eqref{eq: prefactor a diffusion equation monomer conservation} and we can evaluate $g(n,r)$ at the bead radius $R$ to finally derive the local monomer density,
\begin{align}
\begin{aligned}
    g(n,R) &= -\frac{a}{R} + b  \\
    &= \frac{-G(n) R} {V\left[ \frac{4 D}{ k_{\text{poly}} n} + R \left( 1 - \Delta \right)  \right]} +  \frac{G(n)}{V} + \frac{a \Delta}{R} \\
    &= \frac{G(n)}{V \left[ 1 + \frac{R k_{\text{poly}} n}{4 D}\left( 1 - \Delta \right) \right]}.
\end{aligned}
\end{align}
This is the expression given in Eq. \eqref{eq: local actin concentration exact}.

\section{Numerical implementation of spatial diffusion model}
\label{sec: Numerical implementation of spatial model}
We employ the finite element method (FEM) and use the discretization module DUNE-FEM \cite{dednerGenericInterfaceParallel2010} of the DUNE software framework \cite{bastianDuneFrameworkBasic2021} to numerically solve the spatial diffusion model of the actin monomer concentration $g$. The problem is defined by Eqs. \eqref{eq: spatial R-BC}-\eqref{eq: time-dependent F-actin}. To do so, we first derive the weak formulation of Eq. \eqref{eq: time-depndent diffusion} by multiplying with a test function $v$, integrating over the whole domain $V$ and performing partial integration of the Laplace-term with the divergence theorem,
\begin{align}
\begin{aligned} \label{eq: weak form}
    \partial_t \int_V g v \ \text{d}V 
    &= D \int_{\partial V} \mathbf{n} \cdot (\nabla g) v \ \text{d} A - D \int_V \nabla g \cdot \nabla v \ \text{d} V \\
    &\quad + \frac{1}{V} \sum_i^N \xi_i \tilde{k}_{\text{sev}} F_i \int_V v \ \text{d} V \\
    &= - \sum_i^N \frac{k_{\text{poly}} n_i}{4} \int_{\partial \text{Bead}_i} gv \ \text{d} A \\
    &\quad - D \int_V \nabla g \cdot \nabla v \ \text{d} V \\
    &\quad + \sum_i^N \frac{\xi_i \tilde{k}_{\text{sev}} F_i}{V} \int_V v \ \text{d} V.
\end{aligned}
\end{align}
In the last step, we replaced the gradient along the outward-pointing surface normal $\mathbf{n}$ with the Robin boundary condition Eq. \eqref{eq: spatial R-BC} at the inner exclusions. At the outer no-flux boundary, this term vanishes.

For dynamic simulations, we use a backward Euler-scheme to discretize the time derivative in $g$ at time step $m$ as
\begin{equation}
    \partial_t g^{(m+1)} = \frac{g^{(m+1)} - g^{(m)}}{\Delta t}.
\end{equation}
Then, we are able to define
\begin{align}
\begin{aligned}
    a(g^{(m+1)},v) &= \int_V g^{(m+1)} v \ \text{d} V \\
    &\quad + D \Delta t \int_V \nabla g^{(m+1)} \cdot \nabla v \ \text{d} V \\
    &\quad + \sum_i^N \frac{k_{\text{poly}} n_i^{(m)}}{4} \Delta t \int_{\partial \text{Bead}_i} g^{(m+1)} v \ \text{d} A, 
\end{aligned}
\end{align}
and
\begin{equation}
    L^{(m+1)}(v) = \sum_i^N \frac{\xi_i \tilde{k}_{\text{sev}} F_i^{(m)}}{V} \Delta t \int_V v \ \text{d} V + \int_V g^{(m)} v \ \text{d} V.
\end{equation}
We update the network quantities $n_i$ and $F_i$ afterwards by a forward Euler step. Because of this choice, we have reduced the original diffusion Eq. \eqref{eq: time-depndent diffusion} to the linear problem
\begin{equation}
    a(g^{(m+1)},v) =  L^{(m+1)}(v),
\end{equation}
which can be directly handed to the solver in DUNE-FEM. The initial condition must be chosen such that the integral of $g$ over the whole domain equals the total number of actin monomers, $A_{\text{tot}}$. The solution is then given in an appropriate solution space defined on the computational mesh. All meshes, like the one shown in Fig. \ref{fig: spatial 3 beads}(a), have been created with the open-source finite element mesh generator Gmsh \cite{geuzaineGmsh3DFinite2009}.

\section{Model assumptions} \label{sec: Model assumptions}

In our simplified model, we have expressed all necessary quantities in terms of the filament density $n$ within the branching zone in order to keep the final equations to a minimum and allowing for an intuitive understanding.
This was only possible under several assumptions, which we will discuss in this section.

\subsection{Concentrations of actin binding proteins}

As explained in section \ref{sec: Molecular mechanisms underlying network growth}, the creation of new branches is actually a multistep process. There are several pathways with different reaction rates to create a new branch \cite{beltznerPathwayActinFilament2008}. In such a reaction chain with intermediate steps, the relation between the final product and the reagents might be nonlinear.
In principle, the amount of all proteins is finite such that they can become depleted and their concentration could spatially vary.
In Eq. \eqref{eq: filament density ODE}, we have only considered the concentration of actin as a variable and used an effective rate $k_{\text{branch}}$ to absorb the multistep process. Instead of actin, a depletion of activated Arp2/3 might produce similar qualitative results. On the other hand, a depletion of capping proteins would represent a positive feedback and could increase branching activity.

We can obtain a rough estimate of incorporated Arp2/3 and capping proteins in a single network from the average filament length $l_{\text{filament}} \approx 177.5$ nm, estimated in Eq. \eqref{eq: filament length}. This length corresponds to 64 actin monomers per filament. At $0.5 \ \mu$M of actin being polymerized in the network, the amount of integrated Arp2/3 and capping proteins are $0.5 \ \mu \text{M} / 64 = 7.9$ nM each, corresponding to 9 \% and 20 \%, respectively, of the total amount of Arp2/3 and capping proteins (see table \ref{tab: Parameter values}). If we assume a similar diffusion coefficient as for actin monomers, the local depletion is given by Eq. \eqref{eq: local actin concentration exact}. However, the rate of consumption is only $1/64$ compared to the one of actin, amounting to only 2 $\%$ local depletion. Even though Arp2/3 has roughly twice the size of actin monomers and diffusion is probably slower, local depletion would still be small. Thus, it is negligible and the slight global depletion would affect all networks the same way such that it would not change our results qualitatively.

Another possibility would be the depletion of ADF/cofilin as it has been reported by Manhart et al. \cite{manhartQuantitativeRegulationDynamic2019}. Since ADF/cofilin is involved in disassembly and thus only regulating the global amount of actin monomers $G(n)$ in our model, it would also affect all networks the same way and not changing anything fundamentally.

\subsection{Mechanical feedback of the bead} \label{sec: Relevance of the bead}

In our theory, we have neglected the effect of the bead on the network, as if the network could grow freely in the branching zone. We will show now that the mechanical feedback is negligible. If we assume a single filament behaves like a Brownian ratchet \cite{peskinCellularMotionsThermal1993}, the free polymerization velocity, $v_0 = k_{\text{poly}} g(n) d_0$, decreases exponentially under an opposing force $f$,
\begin{equation} \label{eq: polymerization load dependence}
    v_{\text{poly}}(f) = v_0 \exp \left(-\frac{f d_0}{k_{\text{B}}T} \right),
\end{equation}
with the thermal energy $k_{\text{B}}T \approx 4.2$ pN nm at body temperature. For the load-free case, $f=0$, we recover the expression used in Eq. \eqref{eq: polymerization velocity}. To estimate the force per filament we assume that the bead moves at the speed of polymerization and experiences a resistive force based on Stokes' law such that the load per filament is
\begin{equation}
    f = \frac{F_{\text{Stokes}}}{N} = \frac{6 \pi \mu R v_{\text{poly}}(f)}{N},
\end{equation}
with the viscosity $\mu$ of the surrounding medium. Together with Eq. \eqref{eq: polymerization load dependence}, this forms a transcendental equation. But we can evaluate it for typical velocities in bead motility assays or in our model, $v \sim 3 \ \mu \text{m/min}$, and find that the total Stokes friction is $ \sim 2 \ \text{fN} \cdot \frac{\mu}{\mu_{\text{Water}}}$. Therefore, only at a viscosity 1000-fold the one of water, we would enter the regime of the thermal energy and thus have relevant force feedback. However, in our model the force is shared between typically $N \sim 20.000$ filaments growing in parallel and thus the exponent in Eq. \eqref{eq: polymerization load dependence} is basically zero and polymerization happens at the free rate. Because without relevant force feedback the angle between filament and bead surface does not matter for the polymerization rate, we neglect the curved geometry and assume a flat branching zone.

The above considerations coincide with the observation that the bead velocity is independent of the bead radius and of the medium's viscosity up to a resulting drag of several tens of pN, exceeding the viscosity in most bead motility assays by a factor of $10^5$ \cite{wiesnerBiomimeticMotilityAssay2003}. It is important to note that the situation might be very different when the actin network grows on a micropattern on a fixed substrate. Then, polymerization pushes the network backwards, corresponding to a significantly larger resistive force. This might be one of the reasons why Guérin et al. obtain quantitatively different results on micropatterns compared to the bead motility assay.

\subsection{Mesh size of the network} \label{sec: Mesh size of the network}

To derive the depletion of the actin pool, we needed an estimate for the network density along its length and for the length itself. Both quantities depend on the mesh size because we considered filament severing due to the presence of ADF/cofilin. Because actin networks grow in a directed manner they are anisotropic and two different mesh sizes exist: on the one hand, the filament spacing in the cross-section orthogonal to the direction of growth, which corresponds to the definition in Eq. \eqref{eq: filament spacing} in our model. On the other hand, the typical filament length, which we derived in Eq. \eqref{eq: filament length}, provides a measure along the direction of growth. For us there is no fundamental reason to decide which one we should use, especially because Michalski and Carlsson \cite{michalskiModelActinComet2011}, whose ideas we followed to determine the length by the percolation threshold, interpret their lattice model as a coarse-grained representation of an actin network.

However, if we would use the expression of the typical filament in Eq. \eqref{eq: filament length} and insert it together with Eq. \eqref{eq: polymerization velocity} into Eq. \eqref{eq: network length} of the network length, we would obtain
\begin{equation}
    L = \frac{\phi [ \text{cap}] k_{\text{cap}}}{k_{\text{sev}}} \log \left(\frac{1}{p_{\text{c}}} \right).    
\end{equation}
This means the network length would become independent of $n$, while a dependence is observed experimentally \cite{guerinBalancingLimitedResources2025}. For this reason, we have decided to use the filament spacing $1/\sqrt{n}$ as the mesh size.

\subsection{Relevance of fluctuations} \label{sec: Relevance of fluctuations}

In the context of size control, one often considers stochastic fluctuations leading to probability distributions, e.g. of the network density and length.
However, it is not easy to determine where and how noise would enter the equations in our continuum description. A master equation approach on the single filament level is not feasible for a network with several thousand actin filaments growing in parallel. Employing a master equation based on the level of the whole network, i.e. in Eq. \eqref{eq: filament density ODE}, would correspond to the analysis done by Banerjee and Banerjee \cite{banerjeeSizeRegulationMultiple2022}, where they showed that these equations lead to robust size control and coexistence.
But because of this large number of filaments, the length fluctuations experienced by a single filament are actually less relevant and the overall elongation speed of the network will be almost constant. The biggest source of fluctuations is then the fragmentation process at the end of the network. It is not really possible to capture this in continuum models and one would have to employ lattice models. 
However, in our model the disassembly only affects the pool of available monomers. Because larger pieces, which have broken off, must further depolymerize to be actually available for polymerization, the fluctuations in polymerizable monomers due to fragmentation would be buffered.

As discussed in section \ref{sec: Discussion}, the spatial positions of the beads effect the results as well. We cannot account for this in our simple approach encoding everything in filament density $n$. In the spatial model, we could perform simulations for a certain number of beads with different positional arrangements. Translated to the simple approach, this would smear out the sharp transition line in Fig. \ref{fig: Transition from coexistence to selection}(b) and could be interpreted in a stochastic sense of how probable it is for the weak network to form at a random position.

%


\end{document}